\documentstyle{mn}

\newif\ifAMStwofonts

\newcommand{\beq}{\begin{equation}}
\newcommand{\eeq}{\end{equation}}

\ifoldfss
  \ifCUPmtlplainloaded \else
    \NewTextAlphabet{textbfit} {cmbxti10} {}
    \NewTextAlphabet{textbfss} {cmssbx10} {}
    \NewMathAlphabet{mathbfit} {cmbxti10} {} 
    \NewMathAlphabet{mathbfss} {cmssbx10} {} 
  \fi
  \ifAMStwofonts
    \ifCUPmtlplainloaded \else
      \NewSymbolFont{upmath} {eurm10}
      \NewSymbolFont{AMSa} {msam10}
      \NewMathSymbol{\upi}     {0}{upmath}{19}
      \NewMathSymbol{\umu}     {0}{upmath}{16}
      \NewMathSymbol{\upartial}{0}{upmath}{40}
      \NewMathSymbol{\leqslant}{3}{AMSa}{36}
      \NewMathSymbol{\geqslant}{3}{AMSa}{3E}

      \let\leq=\leqslant \let\le=\leqslant
      \let\geq=\geqslant \let\ge=\geqslant
    \fi
  \fi
\fi 

\ifnfssone
  \newmathalphabet{\mathit}
  \addtoversion{normal}{\mathit}{cmr}{m}{it}
  \addtoversion{bold}{\mathit}{cmr}{bx}{it}
  \newmathalphabet{\mathbfit} 
  \addtoversion{normal}{\mathbfit}{cmr}{bx}{it}
  \addtoversion{bold}{\mathbfit}{cmr}{bx}{it}
  \newmathalphabet{\mathbfss} 
  \addtoversion{normal}{\mathbfss}{cmss}{bx}{n}
  \addtoversion{bold}{\mathbfss}{cmss}{bx}{n}
  \ifAMStwofonts
    \ifCUPmtlplainloaded \else
      \UseAMStwoboldmath
      \makeatletter
      \new@mathgroup\upmath@group
      \define@mathgroup\mv@normal\upmath@group{eur}{m}{n}
      \define@mathgroup\mv@bold\upmath@group{eur}{b}{n}
      \edef\UPM{\hexnumber\upmath@group}
      \new@mathgroup\amsa@group
      \define@mathgroup\mv@normal\amsa@group{msa}{m}{n}
      \define@mathgroup\mv@bold\amsa@group{msa}{m}{n}
      \edef\AMSa{\hexnumber\amsa@group}
      \makeatother
      \mathchardef\upi="0\UPM19
      \mathchardef\umu="0\UPM16
      \mathchardef\upartial="0\UPM40
      \mathchardef\leqslant="3\AMSa36
      \mathchardef\geqslant="3\AMSa3E

      \let\leq=\leqslant \let\le=\leqslant
      \let\geq=\geqslant \let\ge=\geqslant
    \fi
  \fi
\fi 

\ifnfsstwo
  \DeclareMathAlphabet{\mathbfit}{OT1}{cmr}{bx}{it}
  \SetMathAlphabet\mathbfit{bold}{OT1}{cmr}{bx}{it}
  \DeclareMathAlphabet{\mathbfss}{OT1}{cmss}{bx}{n}
  \SetMathAlphabet\mathbfss{bold}{OT1}{cmss}{bx}{n}
  \ifAMStwofonts
    \ifCUPmtlplainloaded \else
      \DeclareSymbolFont{UPM}{U}{eur}{m}{n}
      \SetSymbolFont{UPM}{bold}{U}{eur}{b}{n}
      \DeclareSymbolFont{AMSa}{U}{msa}{m}{n}
      \DeclareMathSymbol{\upi}{0}{UPM}{"19}
      \DeclareMathSymbol{\umu}{0}{UPM}{"16}
      \DeclareMathSymbol{\upartial}{0}{UPM}{"40}
      \DeclareMathSymbol{\leqslant}{3}{AMSa}{"36}
      \DeclareMathSymbol{\geqslant}{3}{AMSa}{"3E}

      \let\leq=\leqslant \let\le=\leqslant
      \let\geq=\geqslant \let\ge=\geqslant
    \fi
   \fi
\fi 

\ifCUPmtlplainloaded \else
  \ifAMStwofonts \else 
    \def\upi{\pi}
    \def\umu{\mu}
    \def\upartial{\partial}
  \fi
\fi

\title[Breaking the ``Redshift Deadlock'' - I]{Breaking the 
``Redshift Deadlock'' - I: Constraining the star formation history 
of galaxies with sub-millimetre photometric redshifts}
\author[David H. Hughes {\it et al.}]
{
D.H. Hughes$^{1}$, 
I. Aretxaga$^{1}$, 
E.L. Chapin$^{1}$,
E. Gazta\~{n}aga$^{1,2}$, 
J.S. Dunlop$^{3}$, 
\newauthor
M.J. Devlin$^{4}$, 
M. Halpern$^{5}$, 
J. Gundersen$^{6}$, 
J. Klein$^{4}$,
C.B. Netterfield$^{7}$,
\newauthor
L. Olmi$^{8}$, 
D. Scott$^{9}$, 
G. Tucker$^{10}$
\\
$^{1}$Instituto Nacional de Astrof\'{\i}sica, \'Optica y Electr\'onica 
(INAOE), Aptdo. Postal 51 y 216, Puebla, Mexico \\
$^{2}$Institut d'Estudis Espacials de Catalunya - IEEC/CSIC
Gran Capitan 2-4, 08034  Barcelona, Spain\\
$^{3}$Institute for Astronomy, Univ. of Edinburgh, Blackford Hill, 
Edinburgh, EH9 3HJ, UK \\
$^{4}$Dept. of Physics \& Astronomy, Univ. of Pennslyvania, 
209 South 33rd St., Philadelphia PA 19104-6396, USA\\
$^{5}$Dept. of Physics \& Astronomy, Univ. of British Columbia,
6224 Agricultural Road, Vancouver, B.C. V6T 1Z1, Canada\\
$^{6}$Dept. of Physics, Univ. of Miami,  1320 Campo Sano Drive,
Coral Gables, FL 33416, USA\\  
$^{7}$Dept. of Astronomy, Univ. of Toronto, 60 St. George St.,
Toronto, Ontario, MSS 1A1, Canada\\
$^{8}$Physics Department, University of Puerto Rico, PO box 23343, 
University Station San Juan, PR 00931-3343, Puerto Rico\\
$^{9}$Dept. of Physics \& Astronomy,
Univ. of British Columbia, 2219 Main Mall, Vancouver, B.C. V6T 1Z4, Canada\\
$^{10}$Dept. of Physics, Box 1843, Brown Univ., Providence, RI 02912-1843, 
USA\\
}

\pagerange{\pageref{firstpage}--\pageref{lastpage}}
\pubyear{2001}

\begin{document}

\maketitle 

\label{firstpage}

\begin{abstract}
Future extragalactic sub-millimetre and millimetre surveys have the
potential to provide a sensitive census of the level of obscured star
formation in galaxies at all redshifts. While in general there is good
agreement between the source counts from existing SCUBA (850$\mu$m)
and MAMBO (1.25\,mm) surveys of different depths and areas, it remains
difficult to determine the redshift distribution and bolometric
luminosities of the sub-millimetre and millimetre galaxy
population. This is principally due to the ambiguity in identifying an
individual sub-millimetre source with its optical, IR or radio
counterpart which, in turn, prevents a confident measurement of the
spectroscopic redshift. Additionally, the lack of data measuring the
rest-frame FIR spectral peak of the sub-millimetre galaxies gives rise
to poor constraints on their rest-frame FIR luminosities and star
formation rates. In this paper we describe Monte-Carlo simulations of
ground-based, balloon-borne and satellite sub-mm surveys that
demonstrate how the rest-frame FIR--sub-mm spectral energy
distributions (250--850$\mu$m) can be used to derive photometric
redshifts with an r.m.s accuracy of $\pm 0.4$ over the range $0 < z <
6$. This opportunity to break the redshift deadlock will provide an
estimate of the global star formation history for luminous
optically-obscured galaxies ($L_{\rm FIR} > 3 \times 10^{12}
L_{\odot}$) with an accuracy of $\sim 20$ per cent. 
\end{abstract}


\section{Introduction}
The star-formation history of the high-$z$ starburst galaxy population
can be determined from an accurate measurement of the integral
sub-millimetre (sub-mm) and millimetre (mm) source-counts, and the
luminosity and redshift distributions of sub-mm and mm-selected
galaxies. The contribution of these sources to the total FIR--mm
background, measured by COBE \cite{puget96,hauser98,fixsen98}, places an
additional strong constraint on the acceptable evolutionary models.
It is now generally believed that the luminosity density, due to star
formation activity in galaxies, is roughly constant between $z \sim
1-4$. A series of ground-based 850$\mu$m surveys
\cite{smail97,hughes98,barger98,eales99,eales00,scott01,fox01,borys01},
undertaken with the SCUBA camera \cite{holland99} on the 15-m JCMT and
1.25\,mm surveys \cite{bertoldi00} using the MAMBO camera
\cite{kreysa98} on the 30-m IRAM telescope, have contributed
significantly to this understanding.  Prior to the first SCUBA
results, the optical surveys suggested (contrastingly) that the
density of star-formation declined by a factor of $\sim 5$ over the
same redshift range \cite{steidel96,madau96}. This discrepancy has
highlighted the importance of including a correction for dust
extinction in the optical-UV estimates of star formation
rates. Nevertheless, the contribution of optically-selected starburst
galaxies to the sub-mm background still remains uncertain
\cite{chapman00,peacock00}.  Hereafter, for the sake of brevity,
sub-mm and mm wavelengths will be referred to as {\em sub-mm} unless
explicitly stated otherwise, and a {\em sub-mm galaxy} is considered
to be an optically-obscured starburst galaxy detected in a blank-field
SCUBA or MAMBO survey.

One important caveat to this opening statement is the fact that we
actually have no accurate information on the redshift distribution of
the sub-mm galaxy population. There are $\sim 100$ galaxies identified
in 850$\mu$m surveys at a level $> 2$~mJy. The lack of confident
optical, or IR identifications 
(e.g. Hughes et al. 1998; Richards 1999), and thus
precise redshifts for the sub-mm galaxies, was a source for the early
scepticism of the claims that there existed no evidence for a decline
in density of star formation at $z > 2$.

The use of the radio--sub-mm (850$\mu$m/1.4GHz) spectral index has
gained popularity as a valuable diagnostic of redshift
\cite{hughes98,carilliyun99,carilliyun00,smail00,barger00,dunne00}. The
radio follow-up of SCUBA surveys, and more recently 1.25mm MAMBO
surveys \cite{bertoldi00}, with the VLA at 1.4\,GHz has become the
commonly accepted method to measure the redshift distribution of
sub-mm galaxies. However this technique requires extremely sensitive
observations if radio counterparts are to be detected.  For example,
in order that a radio survey detects $> 90$ per cent of bright sub-mm
galaxies ($S_{850\mu \rm m} > 8$\,mJy) at $z > 1.5$, then a $3\sigma
\leq 7\mu$Jy survey must be conducted at 1.4\,GHz
\cite{aretxaga02}. Unfortunately the necessity of such an ultra-deep
radio survey prohibits this technique from being applied to the
wider-area ($\sim 1-10$~deg$^{2}$) sub-mm surveys that will be
conducted in the next few years (section 2) without an excessive
committment of radio telescope time (e.g. 400--4000 hours per survey). 
An alternative, and, we argue, a more reliable method exists:
namely the use of sub-mm--FIR photometric redshifts, provided that one
can obtain experimental data with sufficient S/N and absolute
calibration accuracy.

\begin{figure*}
\vspace{7cm} \includegraphics{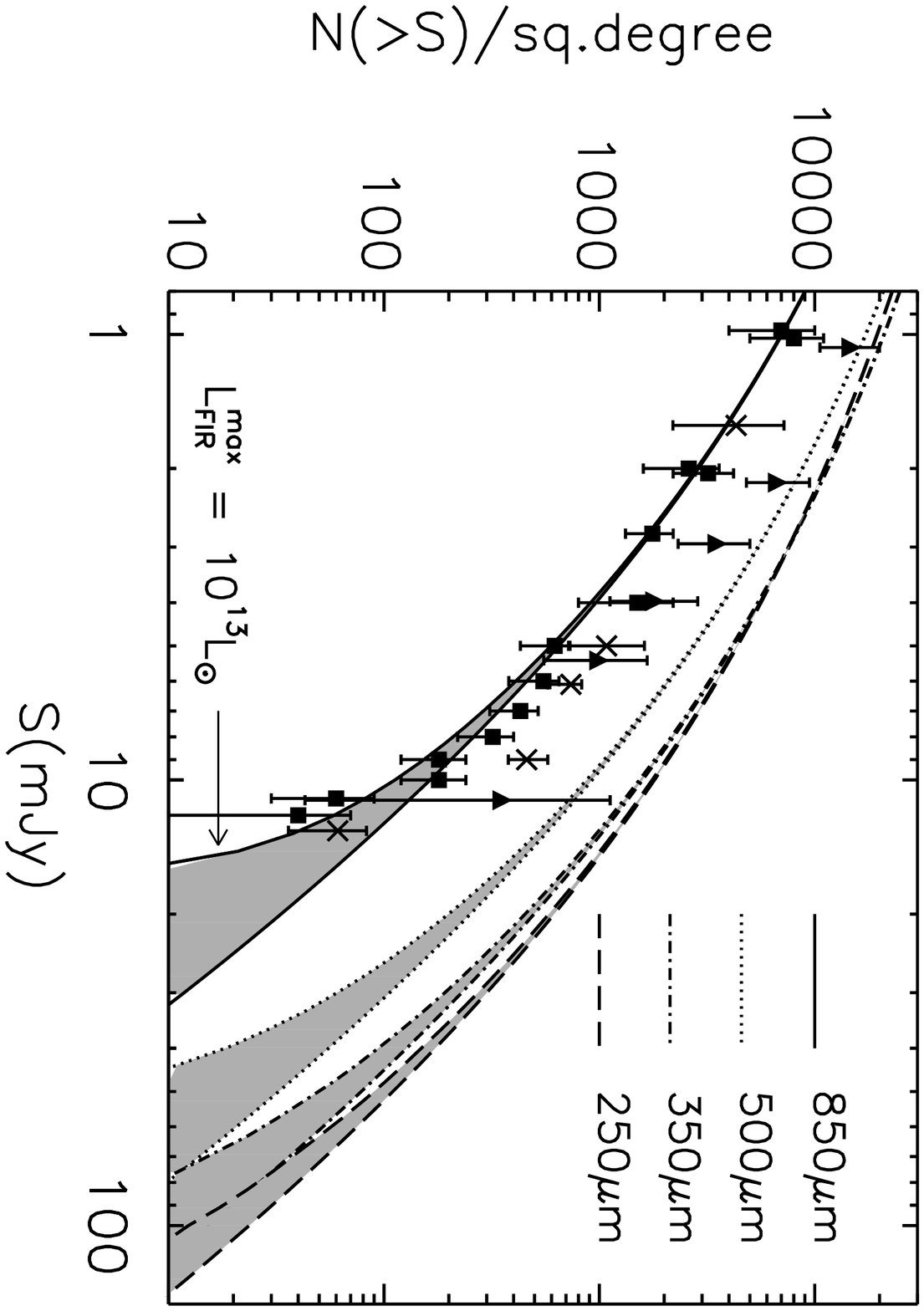} \includegraphics{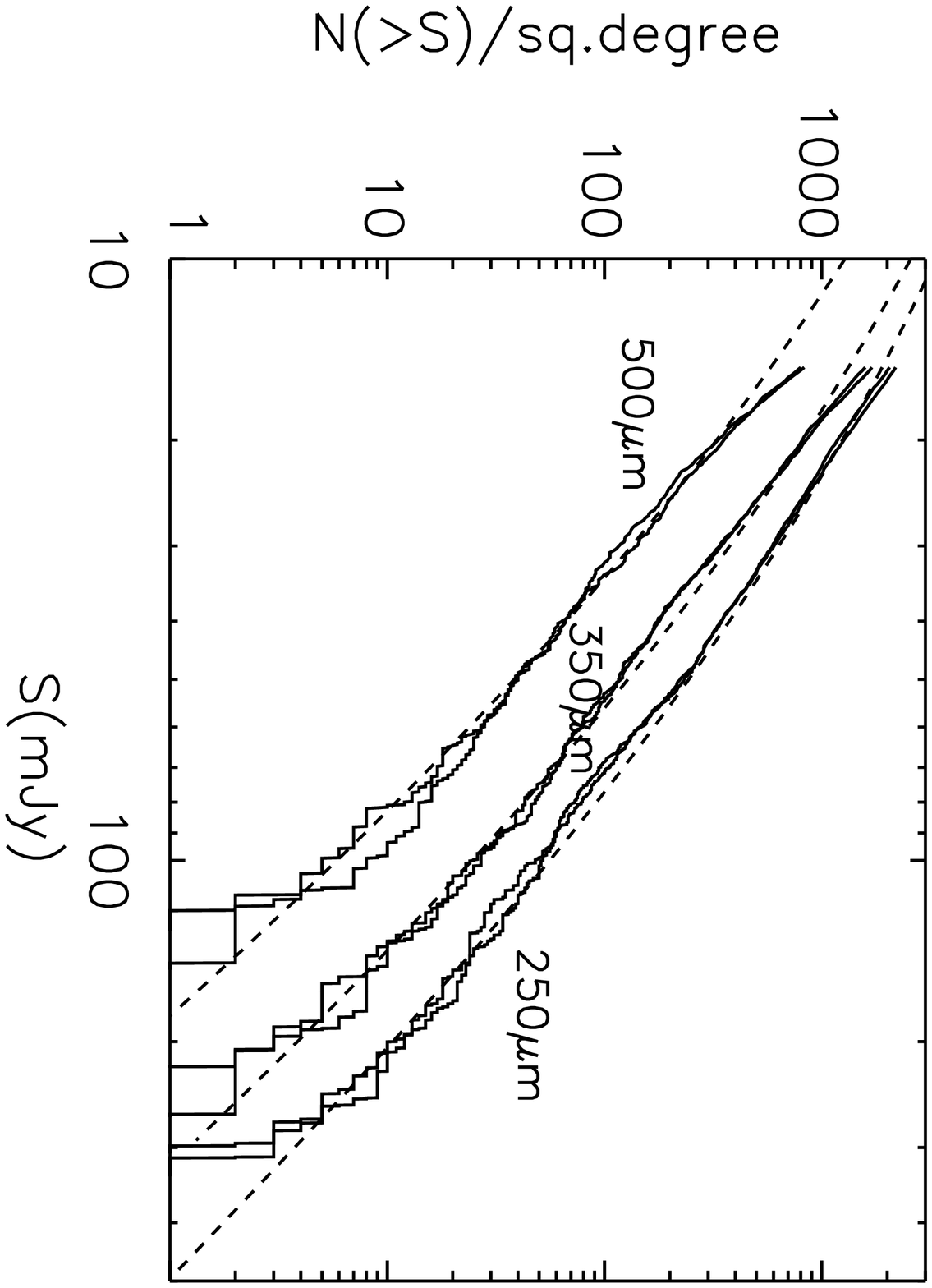}
\caption{Integrated source-counts at sub-mm wavelengths for a model in
which the IRAS 60$\mu$m luminosity function is evolved as
$(1+z)^{3.2}$ upto $z=2$, and maintained constant (with no further
evolution) for $2 < z < 6$. The SED of Arp220 is adopted to represent
the whole galaxy population. {\it Left panel:} The curves represent
the predicted source-counts at 850, 500, 350 and 250$\mu$m. The
measured source-counts from SCUBA surveys at 850$\mu$m (Scott et
al. 2001, and references therein) are shown as filled
squares. Additional 850$\mu$m source-counts from Borys et al. (2001)
are shown as solid triangles.  The flux-densities of the MAMBO 1.25\,mm
source-counts are scaled upwards by a factor of 2.25 to represent the
equivalent measurements at 850$\mu$m, assuming the MAMBO sources lie
at redshifts $0 < z < 6$, and are shown as crosses (Bertoldi et
al. 2000). The models also show the effect of a high-luminosity
cut-off (at $L_{\rm FIR} > 10^{13} L_{\sun}$) in the sub-mm
population. The shaded areas illustrate the regions of parameter space
that need to be searched with future experiments to improve our
understanding of the evolution of the luminous high-$z$ sub-mm
galaxies.  {\it Right panel:} Integrated source-counts for
extragalactic {\em BLAST} surveys at 250, 350 and
500$\mu$m. The dashed-lines show the same evolutionary model described
for the left-panel. The histograms illustrate the extracted
source-counts at each wavelength from two different Monte-Carlo
simulations of 1~deg$^{2}$ surveys that use a library of 13 SEDs. These
simulations explore evolutionary models with small differences in the
strength of evolution and the redshift distribution of sources.  The
details of the Monte-Carlo simulations are summarised in \S 2.2}
\label{fig:counts}
\end{figure*}

Although the possibility of conducting cosmological surveys at sub-mm and mm
wavelengths has been realised in the last few years with 
the development and successful commissioning of sensitive
bolometer arrays, 
the existing 850\,$\mu$m SCUBA and 1.25\,mm MAMBO surveys are limited
in their ability to constrain the evolutionary models of the sub-mm
galaxy population.  The practical reasons for these limitations have
been described elsewhere \cite{hughes00} and can be summarised as
follows: restricted wavelength coverage (enforced by the limited
number of FIR--mm atmospheric windows available to ground-based
observatories); low spatial resolution (resulting in both a high
extragalactic confusion limit and poor positional accuracy);
restricted field-of-view with the current sub-mm and mm bolometer
arrays (typically 5\,sq. arcmin); and low system sensitivity (a
combination of instrument noise, size of telescope aperture and
telescope surface accuracy, sky transmission and sky noise) which
restrict even the widest and shallowest sub-mm surveys to areas $<
0.1$~deg$^{2}$ \cite{scott01,fox01}.  In the effort
to obtain these wide-area shallow surveys, the current sub-mm and mm
observations are necessarily only sensitive to the most luminous
star-forming galaxies ($\rm L_{FIR} > 10^{12} L_{\odot}$, or $\rm SFR
> 100 M_{\odot} yr^{-1}$) assuming the population is dominated by
galaxies at redshifts $> 1$.

The discrepancy shown in Fig.\ref{fig:counts} between the bright-end
source-counts in the 850$\mu$m SCUBA and 1.25\,mm MAMBO surveys
\cite{eales00,bertoldi00,scott01,borys01}, and a visual inspection of
their reconstructed maps, suggests that the clustering of sub-mm
galaxies may be influencing the statistics \cite{egdhh01}.  Taken at
face value, the steepening of the 850$\mu$m counts for sources $> 10$
mJy can be interpreted as evidence for an under-density of galaxies in
the survey-field providing those particular data (due to cosmic
variance), or perhaps a high-luminosity cut-off in the luminosity
function for galaxies $> 10^{13} L_{\odot}$ (Fig.\ref{fig:counts}).  A
plausible alternative explanation is simply the fact that, with so few
bright sources detected (5--10 galaxies in 0.1~deg$^{2}$), the counts
in this regime are extremely sensitive to differences between the SEDs
of the most luminous sub-mm galaxies.  Sensitive, wider-area ($\ge
1$~deg$^{2}$), rest-frame FIR surveys with accurate redshift
information (similar to those discussed in this paper) can distinguish
between these possible alternatives.

\subsection{Ambiguities in the Counterparts of Sub-mm Galaxies}

The current 850$\mu$m SCUBA and 1.25~mm MAMBO surveys 
are struggling to 
identify the sub-mm sources with their optical, IR and radio
counterparts. 
Positional errors of $\sim 2 - 3$ arcsec are associated with the
brightest sub-mm sources ($S_{850\mu m} > 8$\,mJy) which, with some
exceptions, are detected in the widest, shallowest SCUBA surveys
\cite{scott01,eales00}. These errors can be improved with follow-up
mm-interferometric observations
\cite{frayer_agn,gear00,lutz01}. However ambiguous optical
identifications
still remain, even with $\leq 2$ arcsec resolution and sub-arcsec
positional errors (e.g.  Downes et al. 1999).  It should be no
surprise that sub-mm selected galaxies, including those with
mm-interferometric detections, do not always have optical
counterparts, since high-$z$ galaxies observed in the earliest stages
of formation may be heavily obscured by dust. Indeed, this is the most
compelling reason for conducting the sub-mm surveys in the first
instance.  Searches for counterparts at longer IR wavelengths have
proven to be more successful \cite{smail99,frayer00,gear00,lutz01},
and have shown earlier identifications of lower-$z$ bright
optical galaxies to be the incorrect counterpart to the sub-mm galaxies
\cite{smail99,frayer00}.
The natural consequence of these ambiguous and potential
mis-identifications is an inaccurate determination of redshift
distribution, luminosities and star-formation history of high-$z$ galaxies.

\begin{figure*}
\vspace{7.0cm} 
\includegraphics{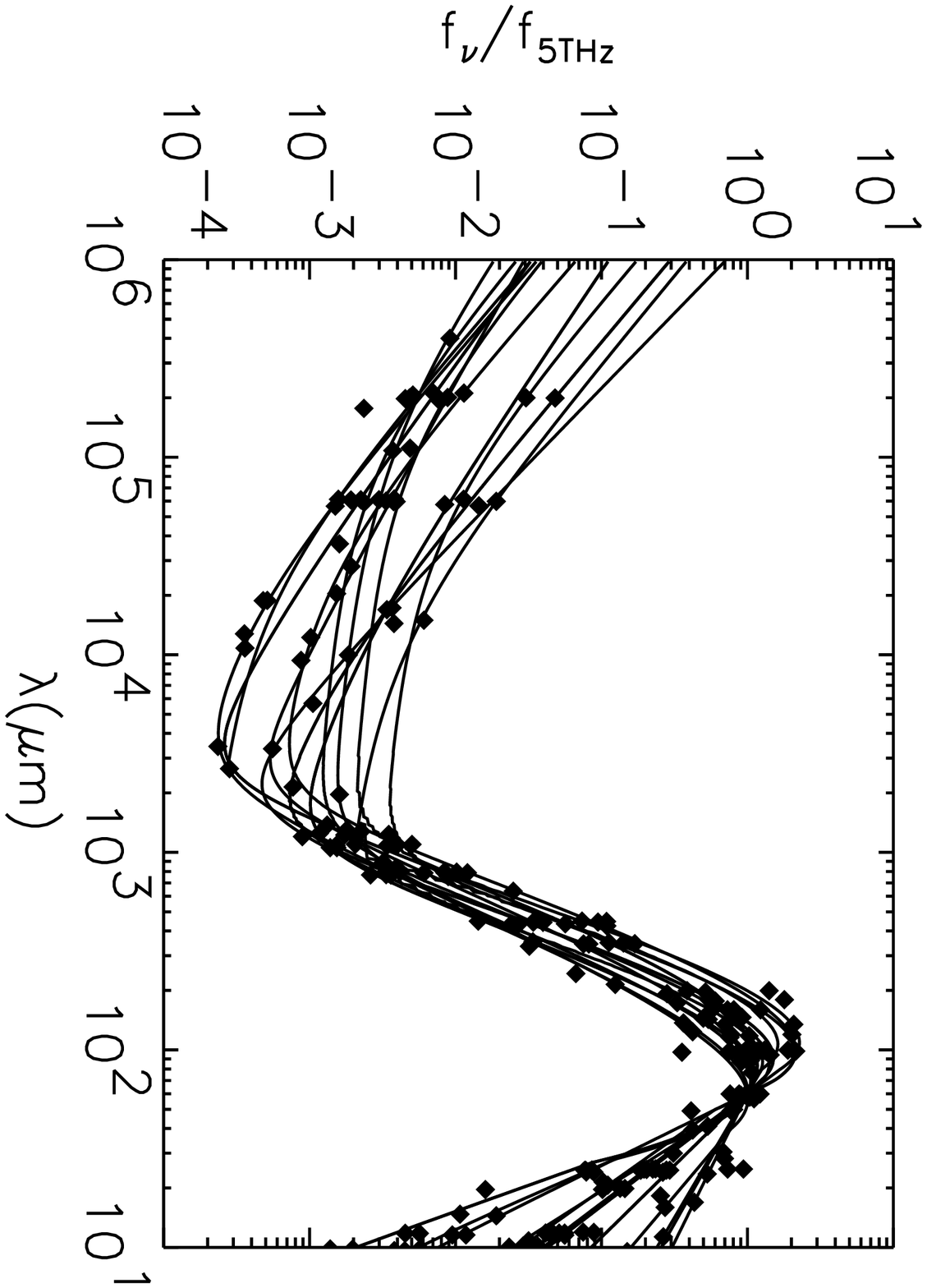} 
\includegraphics{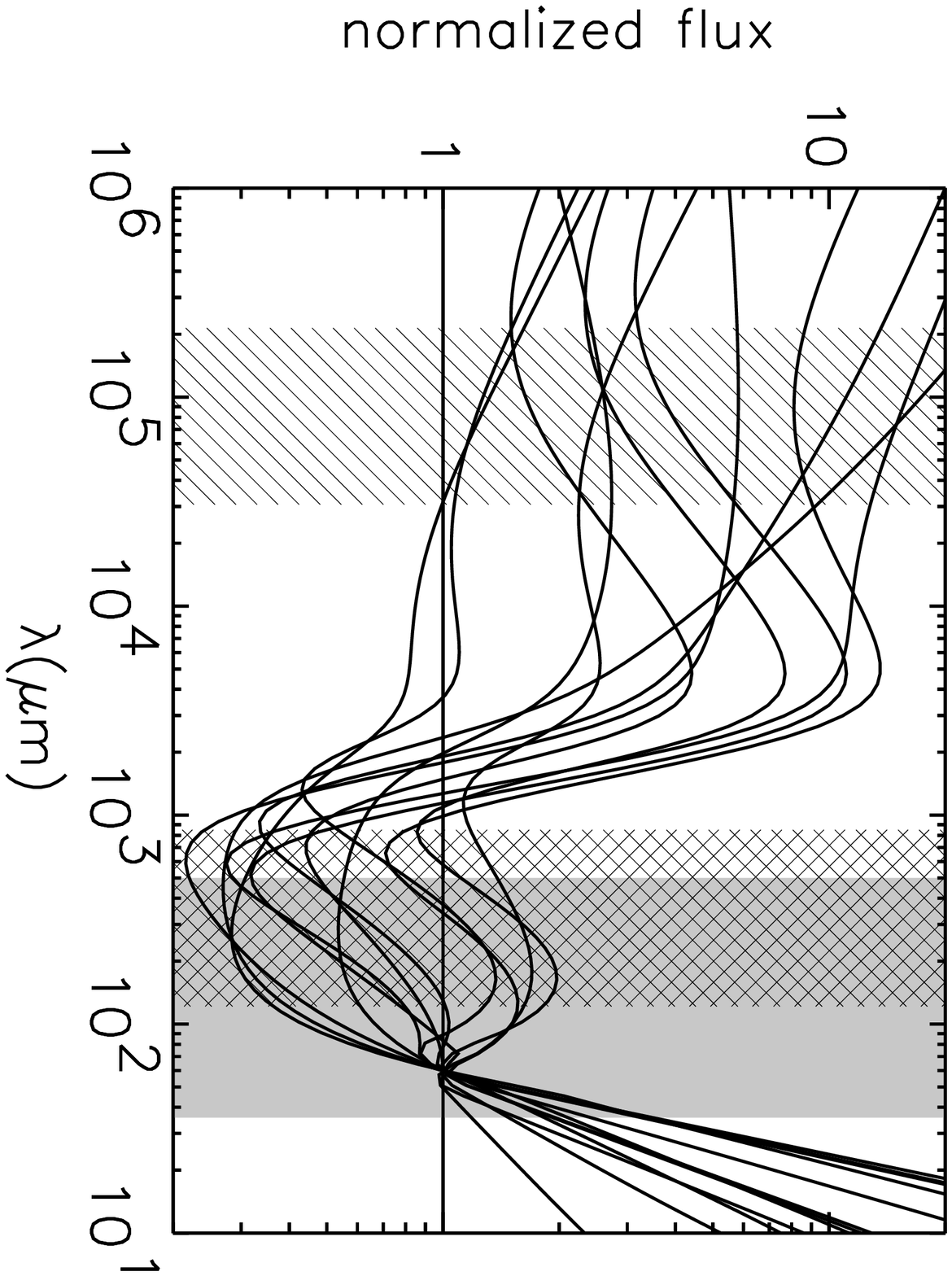} 
\caption{{\it Left panel:} Rest-frame spectral energy distributions
(SEDs) of 13 starburst galaxies, ULIRGs and AGN, normalized at
60$\mu$m.  Lines represent the best-fitting SED models to the data
(diamonds) and include contributions from non-thermal synchrotron
emission, free-free and grey-body thermal emission. The parameters
that describe the composite model for each galaxy, previously
unpublished SED data and references for the remaining SED data are
given in Chapin et al. (in preparation); {\it Right panel:} Composite SED
models for the identical 13 template galaxies shown in the left panel,
normalized at 60$\mu$m and scaled to the flux density of Arp220 at
each wavelength.  The vertically bounded-regions (slanted, crossed and
shaded) correspond to the rest-frame wavelength regimes covered by
observations at 1.4~GHz, 850$\mu$m and in the {\em BLAST} filters
(500--250$\mu$m) respectively for galaxies at $z=0-6$.  This
representation of the SEDs shows more clearly the wavelength regimes
in which the minimum dispersion in galaxy colours can be
expected. Hence photometric redshifts determined from flux ratios
between $\sim 100-1000\mu$m are the most independent of assumptions
regarding the SED of the sub-mm galaxy.}
\label{fig:seds1}
\end{figure*}

\subsection{Uncertainties in the SEDs and Redshifts of Sub-mm Galaxies}

Despite the diversity in the properties of sub-mm galaxies
\cite{ivison_diversity}, efforts have been made to place them in the
context of an evolutionary model of galaxy formation.  The current
estimate of the redshift distribution of sub-mm galaxies is based on
simple comparisons of their IR--radio SEDs with those of redshifted
template SEDs, drawn from local 
starburst galaxies and radio-quiet AGN 
\cite{hughes98,smail99,lilly99,carilliyun99,barger00,carilliyun00,smail00,bertoldi00,dunne00,dunlop01,smail01,chapman02}.
The consensus is that the
population of sub-mm galaxies is distributed between redshifts $z = 1
- 4$, with a median $z = 2-3$. The details of this distribution are
still unknown, and it is common to describe the redshifts of individual
galaxies as ``{\it in the range of ...., with a best guess of ....}''.

This raises an obvious question:{\em What classes of local galaxies
offer the most useful analogues of the high-z sub-mm population?}  In
order to address this we must measure the full rest-frame X-ray to
radio SEDs of individual high-$z$ sub-mm galaxies, measure the
dispersion in the shapes of their SEDs and their range of luminosities
(which may depend on redshift), and thus determine the accuracy with
which the SEDs of high-$z$ sub-mm galaxies can be characterised by a
limited number of local template galaxies.  In practise, the observed
SEDs of high-$z$ sub-mm galaxies are restricted to a few detections
and limits at 850, 450$\mu$m and 1.4\,GHz, with occasional mm
interferometric observations and a K-band identification. No
data exists at rest-frame FIR wavelengths with sufficient sensitivity to 
place additional useful constraints on redshifts.

As mentioned previously, optical and IR follow-up observations of
those few SCUBA sources for which unambiguous identifications exist
have revealed that the sub-mm counterparts are often
optically-obscured, extremely red objects (EROs) with $I-K > 6$
\cite{smail99,gear00,lutz01}, although in a few cases the sub-mm
counterparts are blue galaxies, with a  weak AGN (e.g.
SMM02399--0136, Ivison et al. 2001). Also the suggestion 
that SCUBA galaxies are forming stars at a very high-rate ($> 100
M_{\odot}$/yr) implies that local dusty ULIRGs may also be suitable
analogues of high-$z$ sub-mm sources. Unfortunately the SEDs of this
local class of luminous star-forming galaxies vary significantly with
increasing FIR luminosity \cite{sanders96}. Furthermore, there are few
starburst galaxies at $z > 1$ for which complete IR--radio SEDs exist.

\begin{figure*}
\vspace{22cm} 
\includegraphics{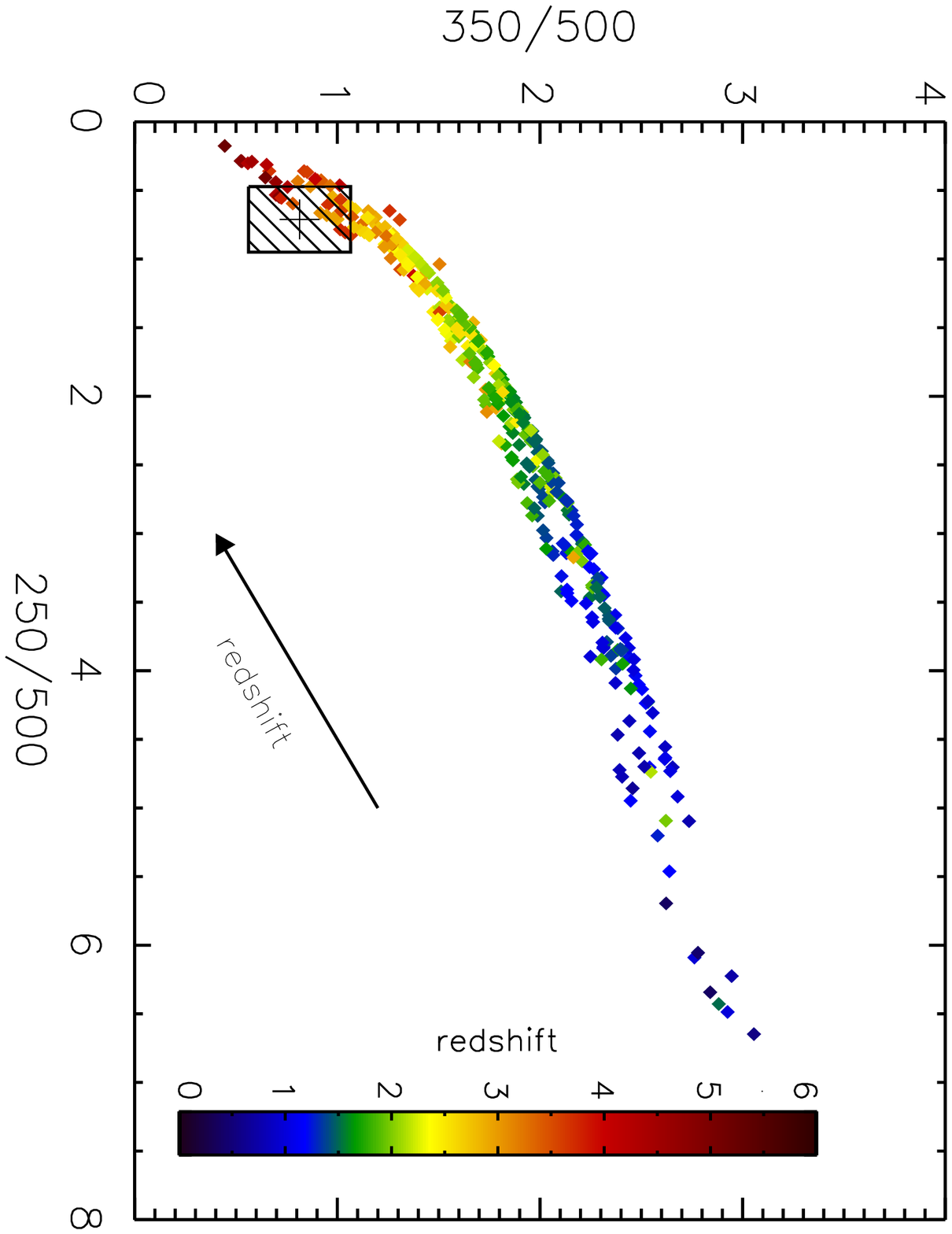} 
\includegraphics{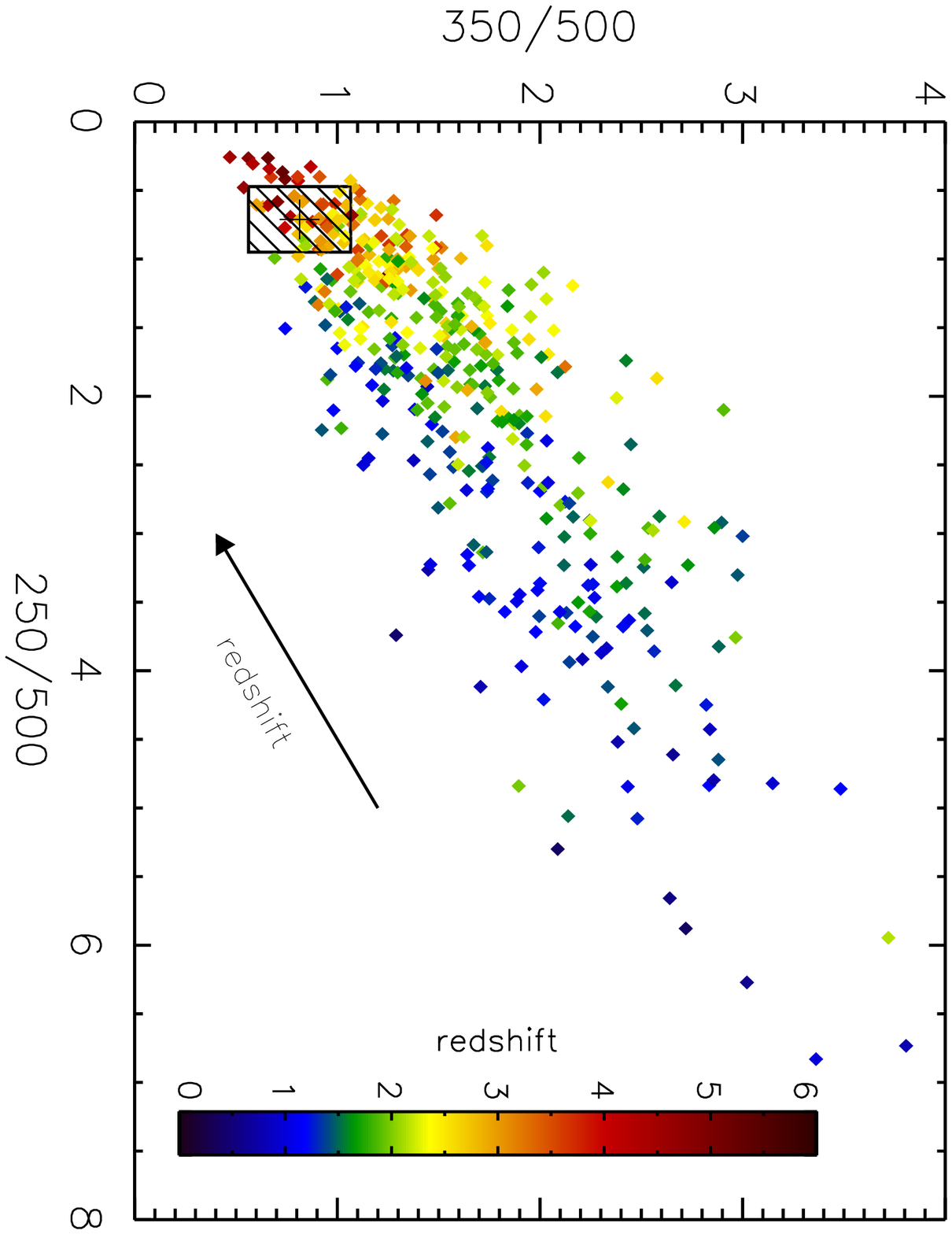}
\caption{Colour-colour (350$\mu$m/500$\mu$m vs. 250$\mu$m/500$\mu$m)
vs. redshift distribution for 424 galaxies detected ($> 3 \sigma$)
simultaneously by {\em BLAST} in all 3 filters in a simulated 1~deg$^{2}$ survey.  {\it Top Panel:} Colour distribution of detected
galaxies without the inclusion of observational (photometric or
calibration) errors.  {\it Bottom Panel:} Colour distribution of
detected galaxies with the inclusion of observational errors (5 per cent
absolute calibration and a random measurement error drawn from a
1$\sigma = 5$\,mJy distribution), which dominate the scatter.  The
black cross marks the position of a $4 \times 10^{12} L_{\odot}$
galaxy at $z=2.81$. The hashed rectangle represents the 1$\sigma$
error on the measured colours. The redshift probability distribution
for this galaxy is shown in Fig.\ref{fig:BLASTz}.  }
\label{fig:BLAST_colours}
\end{figure*}

These initial SCUBA and MAMBO observations have demonstrated the
necessity for larger-area surveys, and also shorter-wavelength
(250--500$\mu$m) sub-mm observations.  Although new (and future)
ground-based facilities (100-m GBT, 50-m LMT, SMA) will offer
partial solutions to some of these problems over the next 6 years,
there will still remain ambiguity in the optical, IR and radio
identifications of sub-mm galaxies. The consequence is a lack of
precision in the redshift distribution and rest-frame FIR
luminosities, and thus an inaccurate constraint on the star formation
history of the entire sub-mm population.

Once it begins operation beyond 2008, ALMA will undoubtedly solve many
of these difficulties with its powerful combination of receivers
operating between 350$\mu$m and 3~mm, and high (sub-arcsec) spatial
resolution.  The expected launch of the Herschel satellite in 2008
will also bring the chance to map large areas of the extragalactic sky
at 250--500$\mu$m with high sensitivity.  However, in late 2003
onwards, a series of long-duration ($\sim 15$ day) sub-orbital balloon
flights will be undertaken by {\em BLAST} -- Balloon-borne
Large-Aperture Sub-millimetre Telescope \cite{devlin01}.  {\em BLAST}
has a 2-m primary aperture, and is equipped with large-format
bolometer cameras operating at 250, 350 and 500$\mu$m which are
prototypes of the SPIRE focal-plane cameras for the Herschel
satellite. {\em BLAST} will conduct sub-mm surveys over $\sim 0.5 -
40$\,deg$^{2}$ (Table\,1), and will provide rest-frame FIR--sub-mm
data for $> 5000$ galaxies, with a point-source sensitivity 3 times
greater than is available from even the largest ground-based sub-mm
telescopes.


In this paper (Paper\,I) we describe Monte-Carlo simulations of sub-mm
surveys that demonstrate how the rest-frame FIR--sub-mm SEDs
(250-500$\mu$m) of galaxies can be used to measure their photometric
redshifts with a conservative 1$\sigma$ accuracy of $\Delta z \sim \pm
0.6$ over the redshift range $1 < z < 6$. Thus we can finally break
the current {\em redshift deadlock} that limits our understanding of
the evolution of the sub-mm galaxy population.  From such precision in
the individual galaxy redshifts, we can determine the star formation
history for luminous sub-mm galaxies brighter than $3 \times
10^{12}L_{\odot}$ with an accuracy of $\sim 20$ per cent. We consider
the specific example of future surveys from {\em BLAST}. Furthermore
we illustrate how these estimates can be improved with sub-mm
observations using the more sensitive SPIRE camera on the {\it
Herschel} satellite, and additionally with the inclusion of
longer-wavelength ground-based 850$\mu$m data from SCUBA.  In
paper\,II of this series \cite{aretxaga02} we develop this
photometric-redshift estimation technique, and apply it to the
existing multi-frequency data for the population of $\sim 140$ sub-mm
galaxies detected in SCUBA and MAMBO surveys. We include, for the
first time, an appropriate treatment of the observational errors at
sub-mm to radio wavelengths, and provide an accurate measurement of
the redshift distribution for sub-mm galaxies.  In paper\,III of this
series (Gazta\~naga et al., in preparation) we address the clustering
of sub-mm galaxies, given their redshift distribution derived from
these photometric redshifts.

The following cosmological model is adopted throughout the paper:
$\Omega_{\Lambda} = 0.7, \Omega_{M} = 0.3, \rm H_{0} = 67\,km s^{-1}
Mpc^{-1}$.


\section{Photometric Redshifts from Wide-Area Sub-millimetre Surveys}
\subsection{Designing the Sub-mm Surveys}
To assess the accuracy of the photometric redshifts that can be
determined from rest-frame FIR--sub-mm data, we generate mock
catalogues of sub-mm galaxies from Monte-Carlo simulations that mimic
realistic observing data (\S 2.2).  Thus, we must first have
prior knowledge of the areas and sensitivities of the sub-mm surveys
that will be conducted with {\em BLAST}, SPIRE, or any other existing or
future sub-mm experiment, in order to achieve the scientific goals. We
have simulated realistic surveys to satisfy this requirement.  These
simulations, which include an evolving population of sub-mm galaxies
that are consistent with the observed number-counts
(Fig.\ref{fig:counts}), and a foreground Galactic cirrus component,
are described in more detail elsewhere
\cite{dhheg00,egdhh01,chapin01}. In particular these simualtions help
understand how the analysis of the experimental survey data can be
affected by telescope resolution, extragalactic and Galactic
source-confusion, survey sensitivity and noise, and sampling variances
due to clustering and shot-noise.

\begin{figure}
\vspace{19.5cm} 
\includegraphics{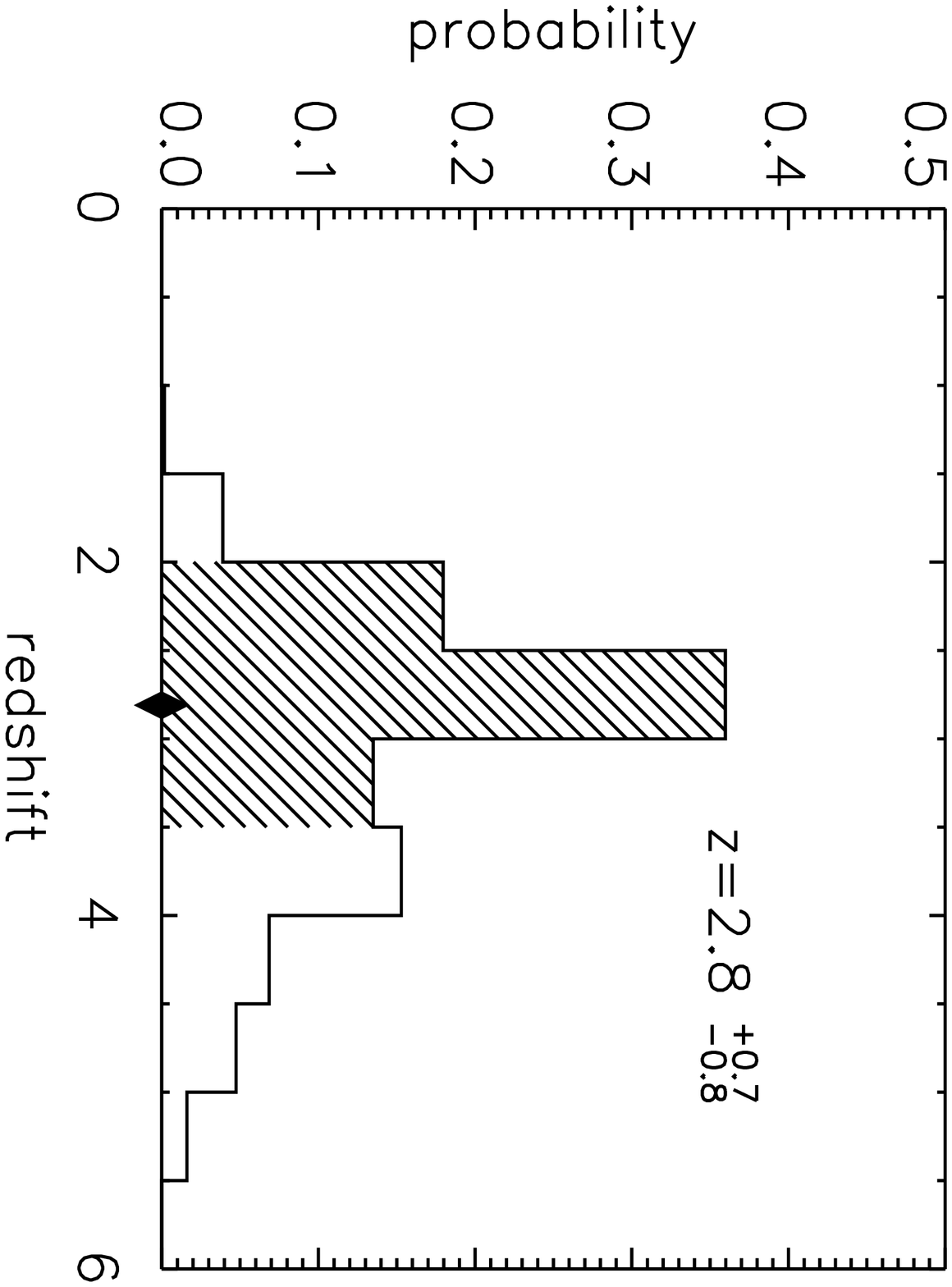}
\includegraphics{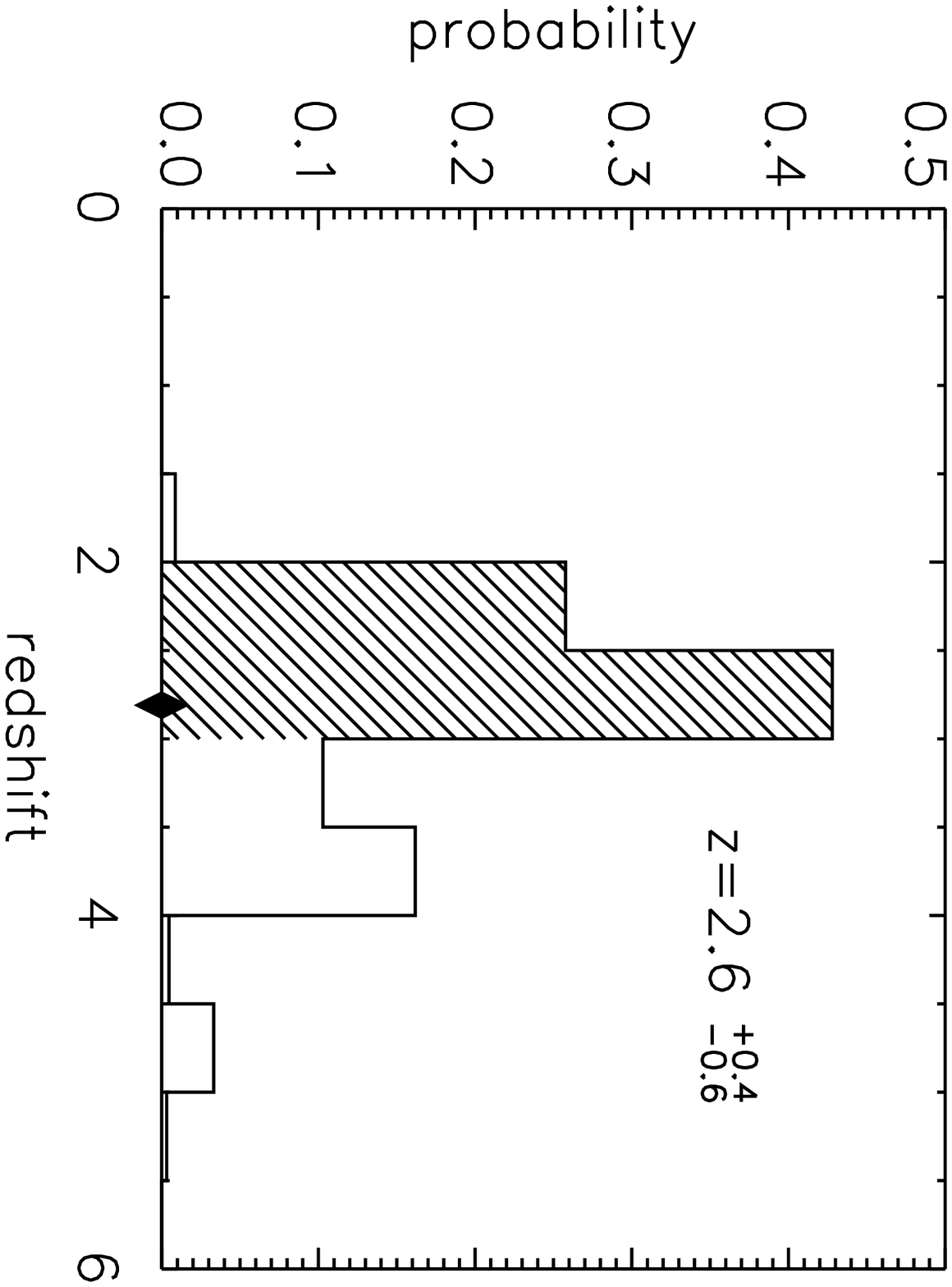}
\includegraphics{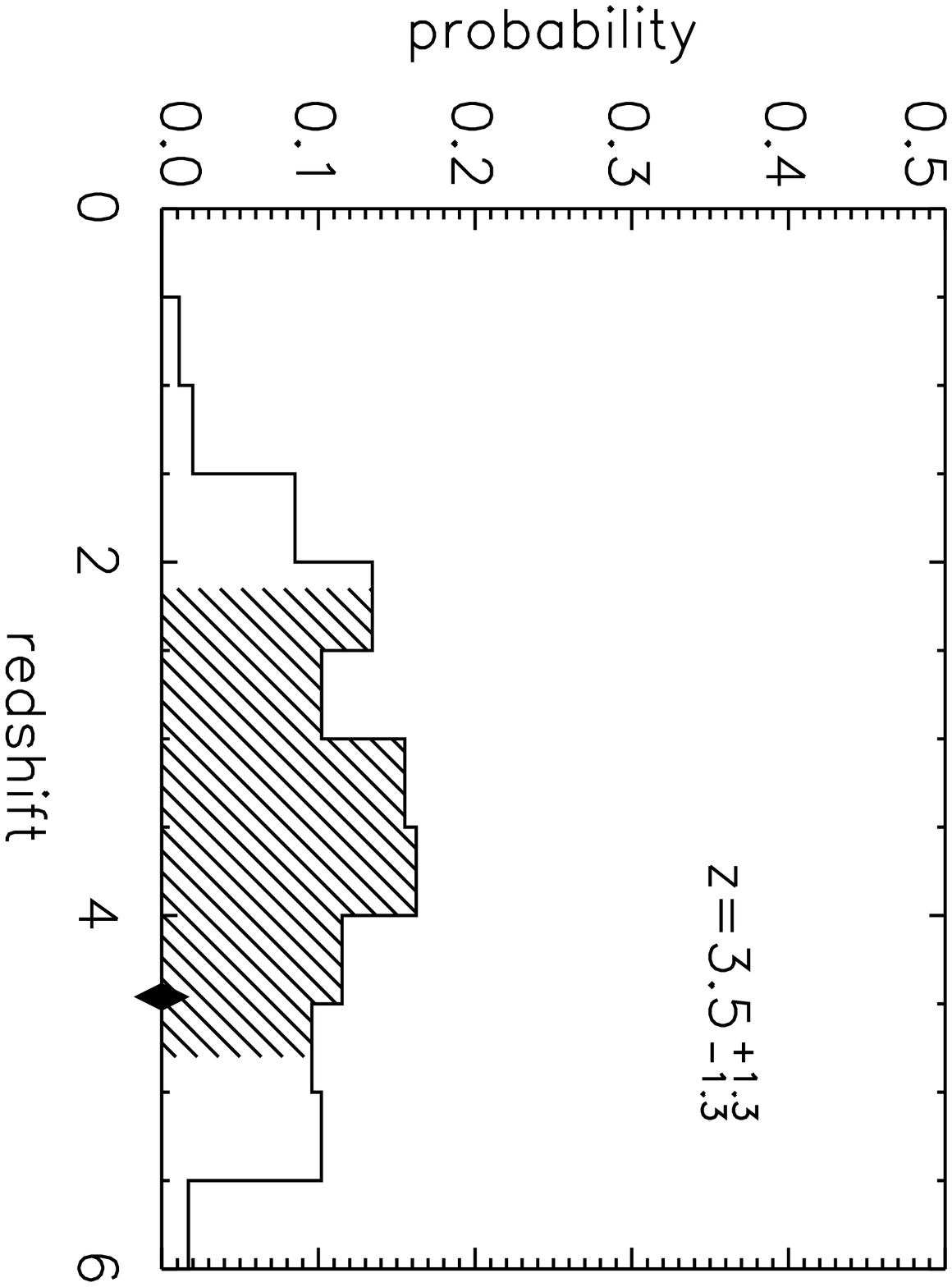}
\caption{Redshift probability distributions for a selection of
galaxies detected by {\em BLAST}. In all panels the diamonds mark the true
redshifts of the galaxies. The photometric redshifts (modes and 68 per cent
confidence intervals) are indicated in the labels and represented as
the shaded areas of the histograms. {\it Top panel:} $4\times 10^{12}
L_{\odot}$ galaxy at $z=2.81$, detected at 250$\mu$m (19\,mJy),
350$\mu$m (21\,mJy) and 500$\mu$m (26\,mJy), slightly below the {\em BLAST}
survey confusion limit. The colours of this galaxy are represented as a
cross in Fig.\ref{fig:BLAST_colours}.  {\it Middle panel:} $6 \times
10^{12} L_{\odot}$ galaxy at $z=2.81$, detected at 250$\mu$m
(34\,mJy), 350$\mu$m (56\,mJy) and 500$\mu$m (45\,mJy) with high S/N, 
above the {\em BLAST} survey confusion limit.  {\it Bottom panel:} $4 \times
10^{12} L_{\odot}$ galaxy at $z=4.46$, detected only at 500$\mu$m
(22\,mJy).}
\label{fig:BLASTz}
\end{figure}

\begin{figure*}
\vspace{11.0cm} 
\includegraphics{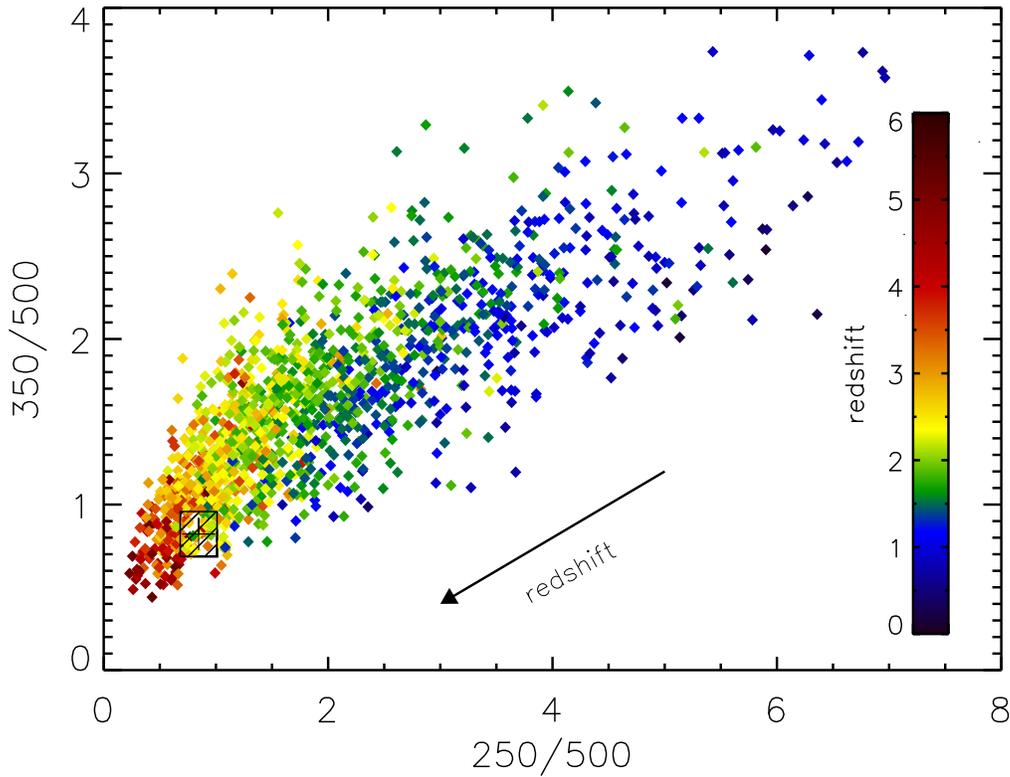}
\caption{Monte-Carlo simulations of the colour--colour (350$\mu$m/500$\mu$m
vs. 250$\mu$m/500$\mu$m) vs. redshift distribution for 1380 galaxies detected
simultaneously by SPIRE at 250, 350 and 500$\mu$m with observational
errors (5 per cent absolute calibration and a random measurement error of
2.5\,mJy) in a 1~deg$^{2}$ survey. The black cross and hashed rectangle
marks the colours for the identical $4 \times 10^{12} L_{\odot}$
galaxy at $z=2.81$ shown in Figs.\ref{fig:BLAST_colours} and
\ref{fig:BLASTz}.}
\label{fig:SPIRE_colours}
\end{figure*}

\begin{figure*}
\vspace{9.cm} 
\includegraphics{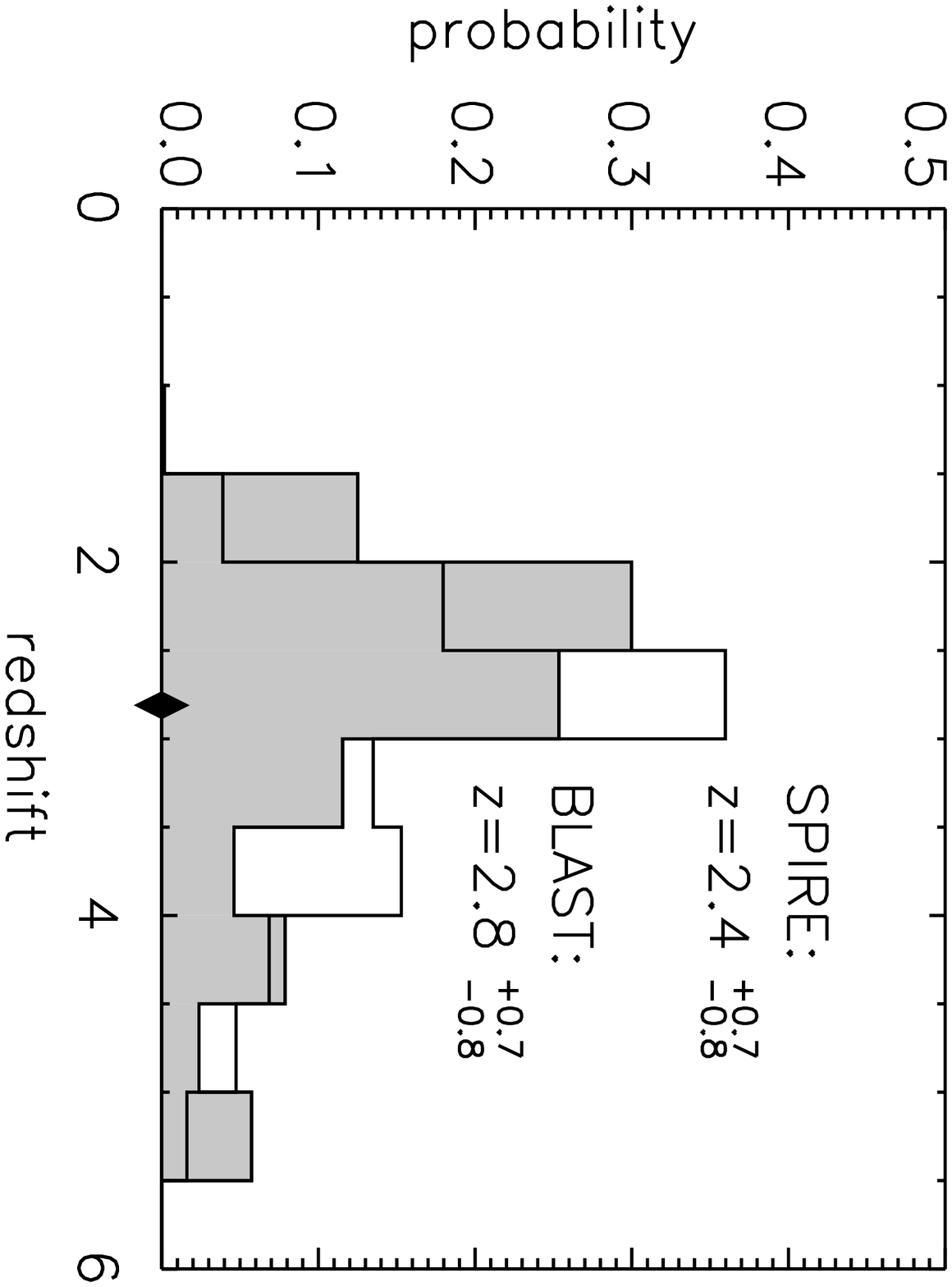}
\includegraphics{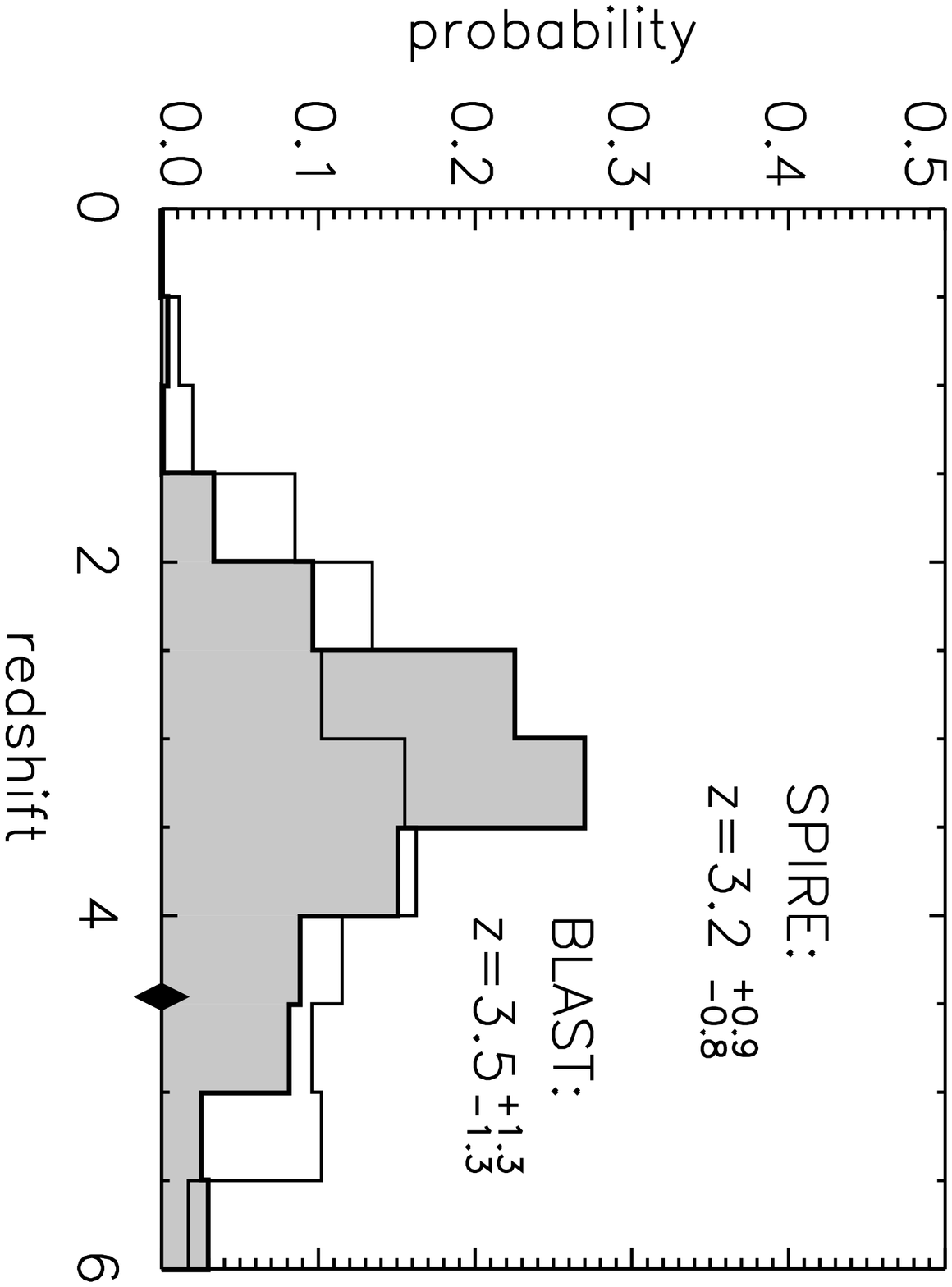}
\caption{A comparison of the redshift probability distributions for
two galaxies observed in a SPIRE survey, and a shallower {\em BLAST} survey
(previously discussed in Fig.\ref{fig:BLASTz}).  Symbols and labels
are the same as in Fig.\ref{fig:BLASTz}.  {\it Left panel:} $4 \times
10^{12} L_{\odot}$ galaxy at $z=2.81$ detected simultaneously with 3
SPIRE filters (shaded histogram), compared with the {\em BLAST} redshift
distribution (open histogram). {\it Right panel:} $4 \times 10^{12}
L_{\odot}$ galaxy at $z=4.46$ detected simultaneously in 2 SPIRE
filters; at 350$\mu$m (11\,mJy) and 500$\mu$m (20\,mJy), compared with
the distribution obtained from a single filter (500$\mu$m) detection
with {\em BLAST}.}
\label{fig:SPIREz}
\end{figure*}

\begin{figure}
\vspace{17cm} 
\includegraphics{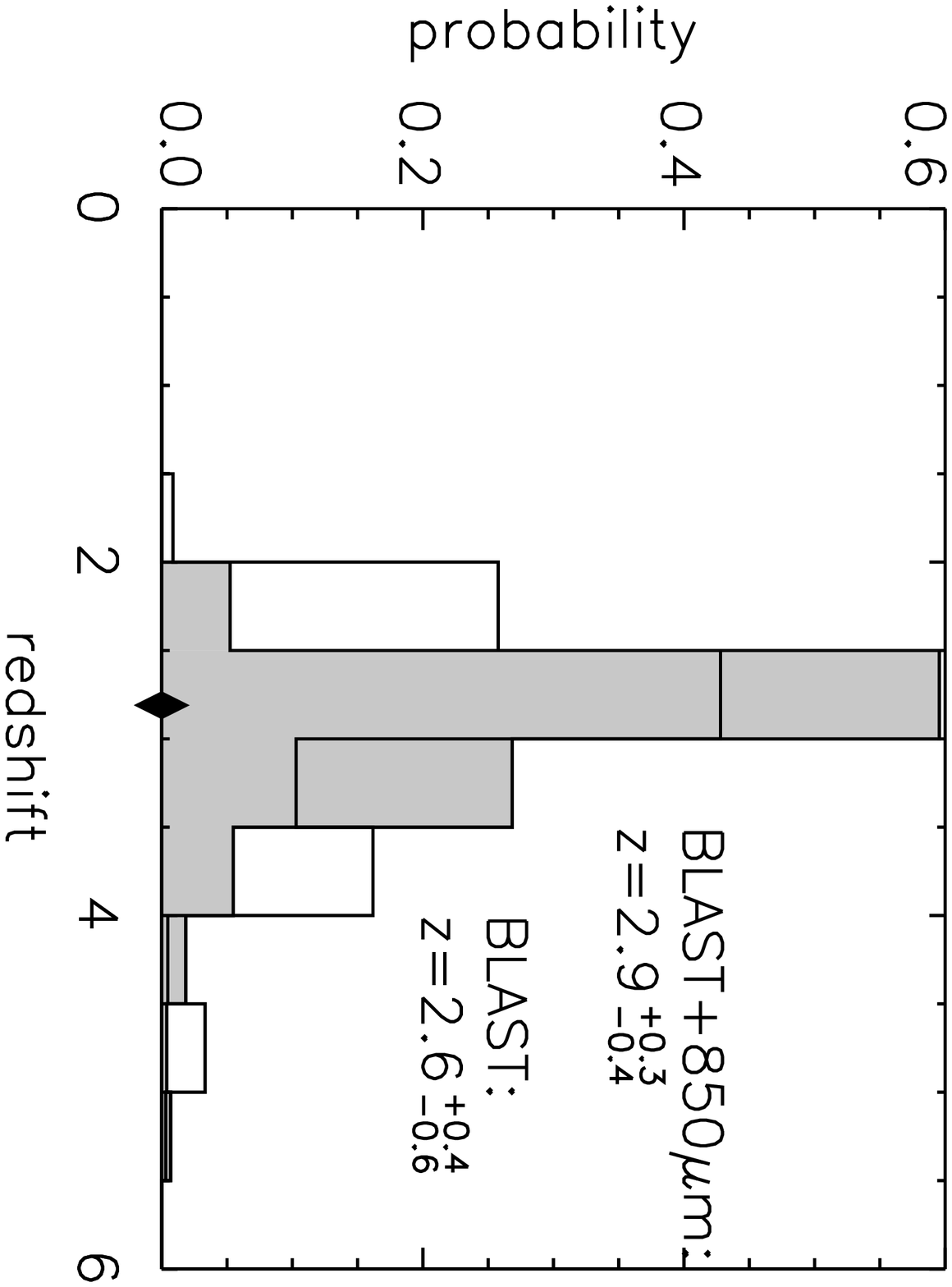} 
\includegraphics{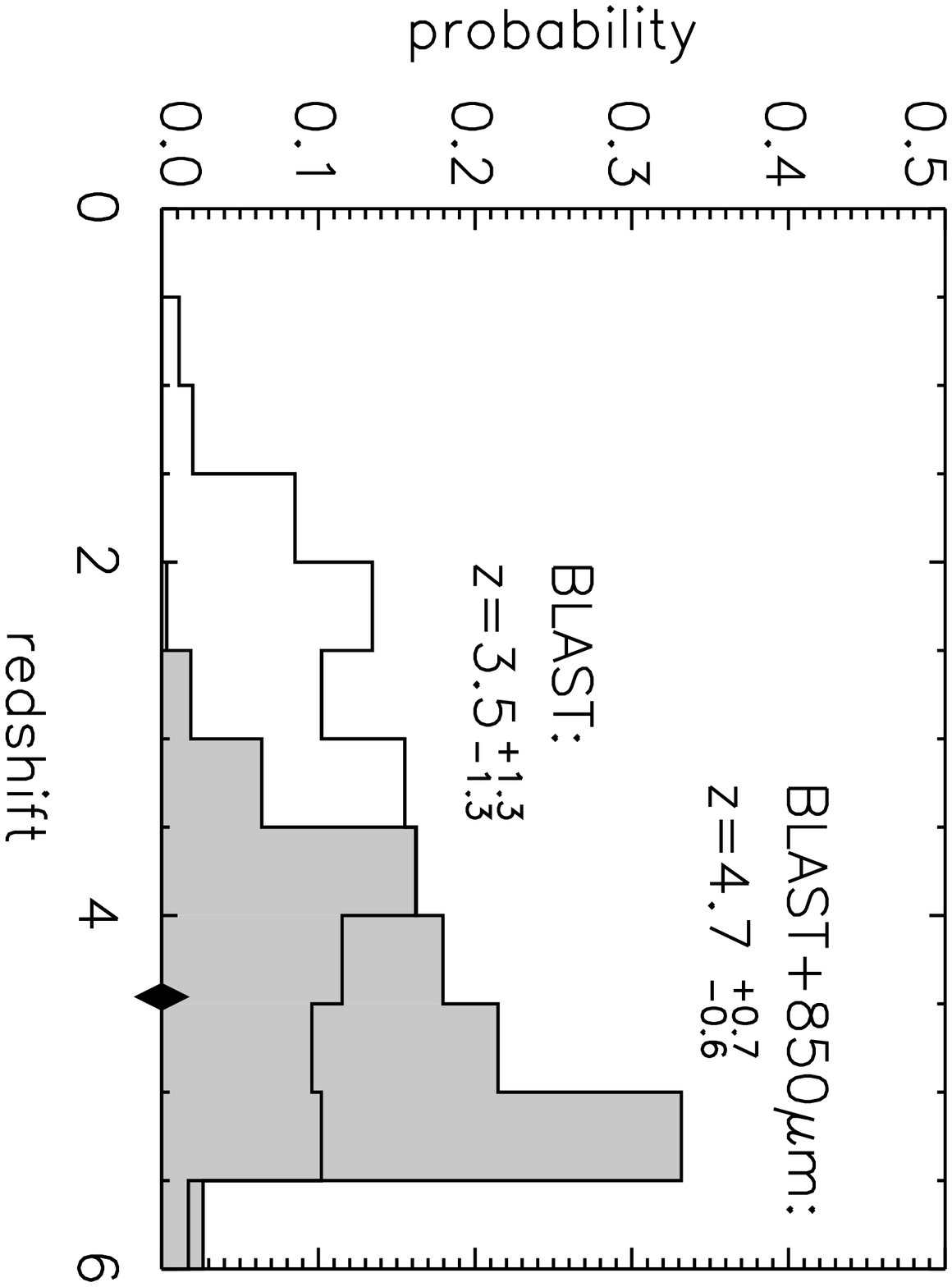} 
\includegraphics{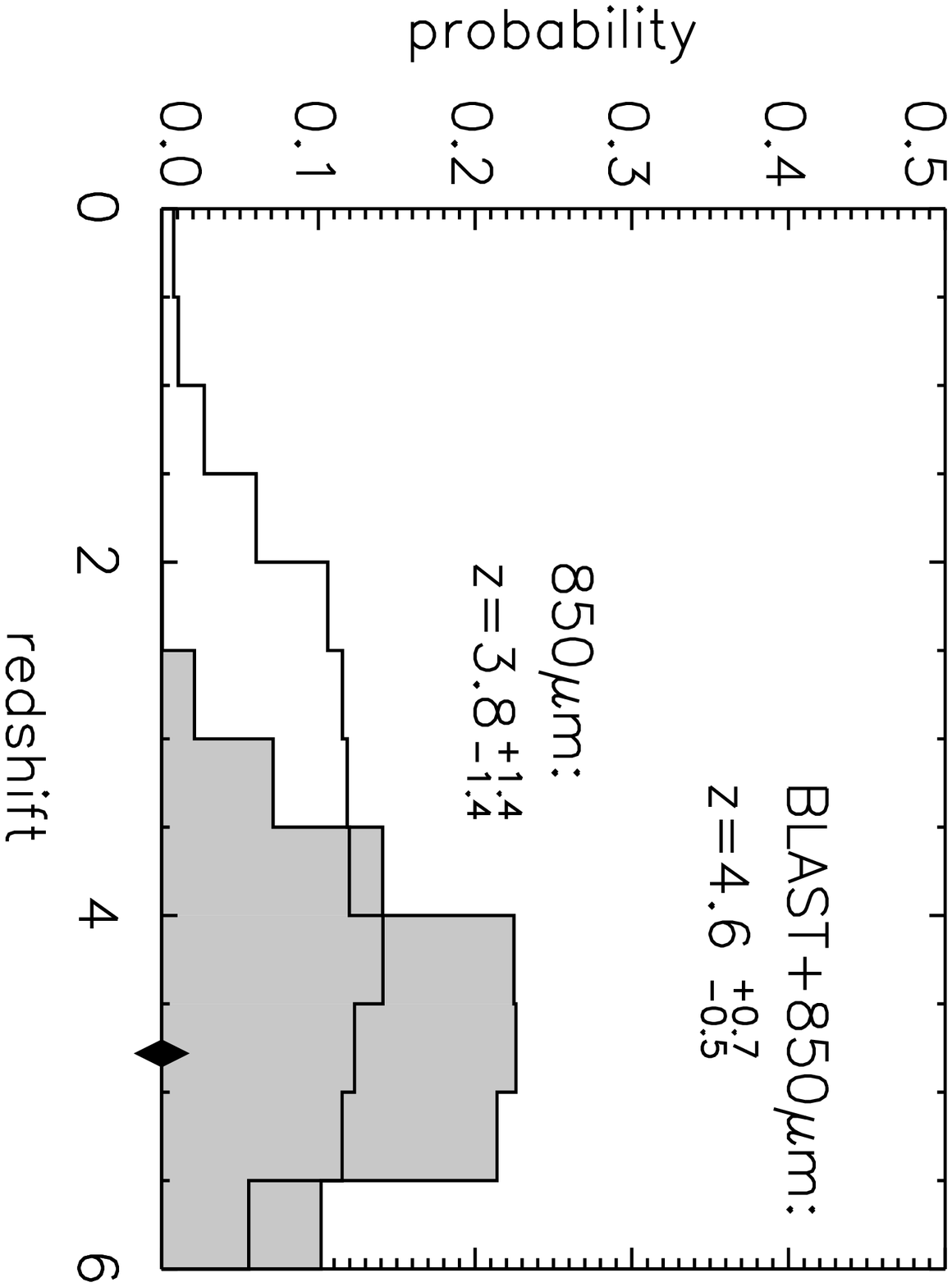}
\caption{Redshift probability distributions demonstrating the
advantages of follow-up observations that extend the sub-mm wavelength
coverage. Symbols and labels are the same as in Fig.\ref{fig:BLASTz}.
{\it Top panel:} $6 \times 10^{12} L_{\odot}$ galaxy at $z=2.81$
detected by {\em BLAST} in all 3 filters (open histogram), previously
discussed in Fig.\ref{fig:BLASTz}.  The redshift estimate improves
with follow-up observations at longer wavelengths.  The shaded
histogram illustrates the benefit of detecting the same galaxy at
850$\mu$m (8\,mJy) in a shallow sub-mm survey ($\sigma_{850\mu \rm m}
= 2.5$\,mJy), e.g.  similar to the UK 8\,mJy SCUBA survey (Scott
et al. 2001); {\it Middle panel:} This example shows the $4 \times
10^{12} L_{\odot}$ galaxy at $z=4.46$, initially detected by {\em
BLAST} in only one filter (500$\mu$m), previously discussed in
Fig.\ref{fig:BLASTz} (open histogram).  Again, the detection at
850$\mu$m (16\,mJy) is sufficiently discriminating to constrain the
distribution to higher redshifts (shaded histogram).  {\it Bottom
Panel:} $5 \times 10^{12} L_{\odot}$ galaxy at $=4.78$ detected at
850$\mu$m (8\,mJy) in a shallow SCUBA survey (open histogram). Despite
the non-detection of this galaxy by {\em BLAST}, in all 3 filters
(250, 350, 500$\mu$m), these shorter wavelength data can place an
improved constraint on the redshift distribution (shaded histogram).}
\label{fig:BLAST_SCUBAz}
\end{figure}

Provided the {\em BLAST} surveys are conducted towards regions of low
Galactic cirrus ($I_{100\mu \rm m} < 1.4$\,MJy/sr), then the {\em
BLAST} survey confusion limit will be dominated by extragalactic
sources. In the absence of any foreground cirrus, the simulation
indicates that the extragalactic $3\sigma$ confusion limit of {\em
BLAST} at 250--500$\mu$m will be $\sim 20-30$~mJy. For comparison the
traditional estimate of extragalactic confusion, 1 source/30 beams
\cite{condon74}, suggests that confusion will begin to dominate below
30~mJy.  

Hence, during a single long-duration balloon flight, {\em BLAST} will
conduct a series of simultaneous large-area surveys ($\sim 0.5 -
40$~deg$^{2}$) at 250, 350 and 500$\mu$m, down to a 1$\sigma$
sensitivity of 5~mJy, detecting in each survey $\sim 1000$ luminous
sub-mm galaxies ($> 10^{12} L_{\odot}$) between $0.5 < z < 6$
(Table~1). The accuracy with which photometric redshifts can be
determined from surveys of this magnitude is discussed in the
following sections.

\begin{table*}
\begin{center}
\begin{tabular}{ccccccc}
\\
\hline 
survey area         & 
1$\sigma$ depth     & 
no. of pixels       & 
\multicolumn{2}{c}{no. of detected galaxies}   & 
\multicolumn{2}{c}{no. of $>5\sigma$ galaxies} \\ 

(deg$^{2}$) & 
              & 
              & 
$> 5  \sigma$  & 
$> 10 \sigma$  & 
$ z > 1$   & 
$ z > 3$   \\
\hline
1.0    & 5  mJy &  18334  & 835  & 265 & 765  & 147 \\
2.0    & 7  mJy &  36668  & 1012  & 291 & 927  & 151 \\ 
4.0  & 10 mJy &  73336  & 1100  & 294 & 988  & 147 \\
9.0   & 15 mJy &  165006 & 1111  & 247 & 1023  & 129 \\
36.0   & 30 mJy &  660024 & 990  & 246 & 895  & 105 \\
\hline
\end{tabular}
\caption{Examples of the number of detected galaxies and their
redshift distributions in alternative 50~hour 250$\mu$m {\em BLAST}
surveys to be undertaken during a series of future long duration
($\sim 10 - 15$ day) balloon flights.
}
\end{center}
\vspace{-.16in}
\label{tab:counts_tab}
\end{table*}

\subsection{Monte-Carlo Simulations of Sub-mm Galaxy Colours}

Measurements of photometric redshifts determined from
FIR--sub-mm--radio data usually do not include a rigorous analysis of
the errors and the uncertainties in the calculations
{\cite{hughes98,blain99,carilliyun99,carilliyun00,dunne00,eales00,barger00}.
The desire to assign a degree of confidence (probability) to the
photometric redshift of any individual sub-mm galaxy motivated the
development of the Monte-Carlo simulations, taking into account the
imprecise knowledge of the luminosity function of sub-mm galaxies, the
dispersion in the luminosity-dependent SEDs, absolute calibration and
observational photometric errors \cite{aretxaga01,aretxaga02}. From
these individual redshifts it is then possible to provide a robust
statistical measure of the redshift distribution for the entire sub-mm
population.

In the case of {\em BLAST} (and also SPIRE), the power of this simple
technique to derive redshifts arises from the unique ability of these
experiments to conduct observations at 250--500$\mu$m that bracket the
ubiquitous rest-frame FIR peak (at $\sim 60-150\mu$m) in the SEDs of
high-redshift ($1 \le z \le 4$) galaxies undergoing a significant
amount of star formation.

The adopted recipe that generates mock catalogues of galaxies between
$z=0$ and $z \sim 6$, for a given wavelength and
under the chosen cosmological model, is summarised:

\begin{enumerate}
\item 
use a single representative SED to calculate the distribution of
rest-frame FIR luminosities of galaxies, at all redshifts, from an
evolving 60$\mu$m local luminosity function, $\phi[L,0]$ \cite{saunders90}, that reproduces the observed sub-mm number
counts. We adopt a model of pure luminosity evolution,
$\phi[L,z]=\phi[L/L^*(z)]$, where

\begin{equation}
 L^*(z)= \left\{
 \begin{array}{l}
     \begin{array}{ll}
      (1+z)^{3.2} \,\, L^*(0)    &  \hspace*{0.5cm} \mbox{for \ \ } z \le 2 \\
      33.6 \,\,L^*(0)     &  \hspace*{0.5cm} \mbox{for \ \ } 2 < z \le 6 \\
      \end{array}
   \end{array}
   \right.
 \label{eqn:evol}
\end{equation}

\noindent It has been verified that an alternative choice of SED can
compensate for small differences in the evolutionary model, and hence still
reproduce the source-counts (Fig.\ref{fig:counts});

\item
randomly assign an SED from a library of 13 template starburst
galaxies, ULIRGs and AGN (Fig.\ref{fig:seds1}) to each of the galaxies
in the mock catalogue, and use this SED to calculate the intrinsic
sub-mm colours.  The most crucial assumption of the paper occurs at
this point in the simulation. We assume no correlation between the
shape of the SED and luminosity (or redshift), and instead rely on the
assumption that our library of 13 template SEDs is representative of
those that exist at high-$z$. The library, which includes M82,
NGC1614, NGC3227, NGC2992, NGC4151, NGC7469, NGC7771, IZw1, Mkn231,
Mkn273, Arp220, IRAS10214 and the ``Cloverleaf'' quasar (H1413+117),
contains galaxies with FIR luminosities in the range of ${\rm
log}\,L_{\rm FIR}/L_{\odot} = 9.6 - 12.3$. The sample includes a
mixture of AGN and starburst galaxies, and composite galaxies with
both an AGN and a luminous extended FIR starburst region. The dust
temperature ($\sim 30 - 60$\,K) and emissivity index ($\beta = 1.4 \pm 0.3$), 
which define the shape of their rest-frame FIR--sub-mm spectra
are independent of the type or strength of nuclear activity.  As
mentioned in \S 1, the sensitivity of the sub-mm observations
restricts the surveys to a detection threshold of ${\rm log}\,L_{\rm
FIR}/L_{\odot} > 12$.  Thus, an implicit assumption in this method is
that one can simply scale the SED of a lower luminosity galaxy to
represent that of a high-redshift sub-mm galaxy.  Furthermore, there
is no conclusive evidence that the rest-frame FIR--sub-mm SEDs of
galaxies vary with luminosity, and redshift beyond the range of SEDs
considered here - there is simply a lack of sufficient high
signal-to-noise observational data, over the appropriate wavelength
regime, to come to any secure conclusions;

\item
add random noise (drawn from the distribution describing the
measurement errors), and add a constant absolute flux calibration error to
the intrinsic fluxes to allow a more realistic assessment of the
photometric redshift accuracy: 1$\sigma$ photometric errors of 5 and
2.5 mJy for the {\em BLAST} and Herschel/SPIRE observations respectively,
and in both cases an absolute calibration error of 5 per cent.  The combined
simulated images of extragalactic sources and cirrus, described
earlier in \S 2.1, determine the depth of the surveys considered here,
ensuring that galaxy confusion dominates. However this Monte-Carlo
treatment of the photometric accuracy does not consider the
contaminating effects of cirrus, clustering of galaxies, shot-noise or
projection effects. These considerations are deferred to a future
paper; 

\item finally, extract flux-limited catalogues from a 1~deg$^{2}$ 
confusion-limited surveys at 250, 350 and 500$\mu$m with $1\sigma =
5$\,mJy and 2.5\,mJy sensitivities, appropriate for {\em BLAST} and SPIRE
observations.

\end{enumerate}

Fig.\ref{fig:BLAST_colours} shows an example of one of the possible
combinations of sub-mm colours, for a catalogue of $\sim 424$ galaxies
detected ($>3\sigma$) simultaneously at 250, 350 and 500$\mu$m in a
1~deg$^{2}$ {\em BLAST} survey. As expected the dispersion in sub-mm
colours is dominated by the observational errors and not by the
differences between the shapes of the template SEDs.  This is
particularly true when the colour ratios are determined over a
relatively small wavelength interval, as is the case for the 250, 350
and 500$\mu$m observations. However, as we extend the wavelength
baseline over which colours are calculated then the effect of
differences in the galaxy SEDs on the photometric colours becomes
increasingly significant (see Fig.\ref{fig:seds1}). An example of this
important effect can be found with the inclusion of longer wavelength
data to the {\em BLAST} observations (\S 2.3.3), or perhaps more
extreme, with the combination of 850$\mu$m and 1.4GHz fluxes
\cite{carilliyun00} which are separated in wavelength by a factor $>
200$.  Thus, in general, without an adequate treatment of
observational errors and a representative set of reference SEDs,
photometric redshifts will be estimated with an over-optimistic
accuracy.

\subsection{Sub-mm Photometric Redshifts}

Sub-mm photometric redshifts can be determined by calculating the
probability that the the colours (including errors) of an observed
sub-mm galaxy are consistent with the colours of every galaxy (with a
known redshift) in the mock catalogue. The final redshift probability
distribution, $P(z)$, of any galaxy 
is then simply the sum of the individual probabilities from the entire
catalogue, or explicitly

\begin{equation}
P(z) = a \sum_{i, \forall z_i} 
\Phi( {\bf c}_i - {\bf c}_0 ) \mbox{\ \ \ ,}
\end{equation}

\noindent
where $a$ is the normalization constant, such that $\int_0^{z_{\rm
max}} P(z)dz =1$ where $z_{\rm max} = 6$.  The multi-variate Gaussian
probability distribution, $\Phi$, for $k$ colours, is given by

\begin{eqnarray}
\Phi({\bf c}_i - {\bf c}_0) = (2\pi)^{-k/2} \mid {\bf A}^{-1} 
\mid^{1/2} \times & \nonumber  \\
& \hspace{-3.4cm} \exp \left( -\frac{1}{2} {\bf (c_i - c_0)' A^{-1} (c_i - c_0) }
\right) 
\mbox{\ \ \ ,}
\end{eqnarray}

\noindent
where ${\bf c}_i$ is the colour vector of the $i$th mock galaxy, such that
$z_i \in [z-dz,z+dz]$, and ${\bf c}_0$ is the colour vector of the test
sub-mm galaxy, for which we want to derive its redshift distribution.
${\bf A}$ is the covariance matrix of the colour distribution, 
of elements

\begin{equation}
A_{\alpha \beta}= \langle (\bar{c}_{\alpha} - {c}_\alpha)
			  (\bar{c}_{\beta} - {c}_\beta)
	          \rangle  \mbox{\ \ \ ,}
\end{equation}

\noindent
for two colours $\alpha$, $\beta$. In the majority of the analysis
performed for this paper, the non-diagonal elements of the covariance
matrix are small. When more than two colours are considered, for
simplicity, the above prescription can be substituted with standard
$\chi^{2}$ statistics.

In this analysis, we have adopted redshift bins of width 0.5 ($dz=0.25$).
The most probable redshift of the test sub-mm galaxy will be given by the 
mode of the discrete $P(z)$ distribution, estimated as the centroid of the Gaussian 
that best fits the 5 bins around the maximum of $P(z)$. The asymmetric
error bars ($z_{-}$, $z_{+}$)
correspond to 68 per cent confidence levels such that 
$\int_{z_{-}}^{z_{+}} P(z)dz =0.68$ and 
$(z_{+} - z_{-})$ is minimised.

\begin{figure*}
\vspace{9cm} 
\includegraphics{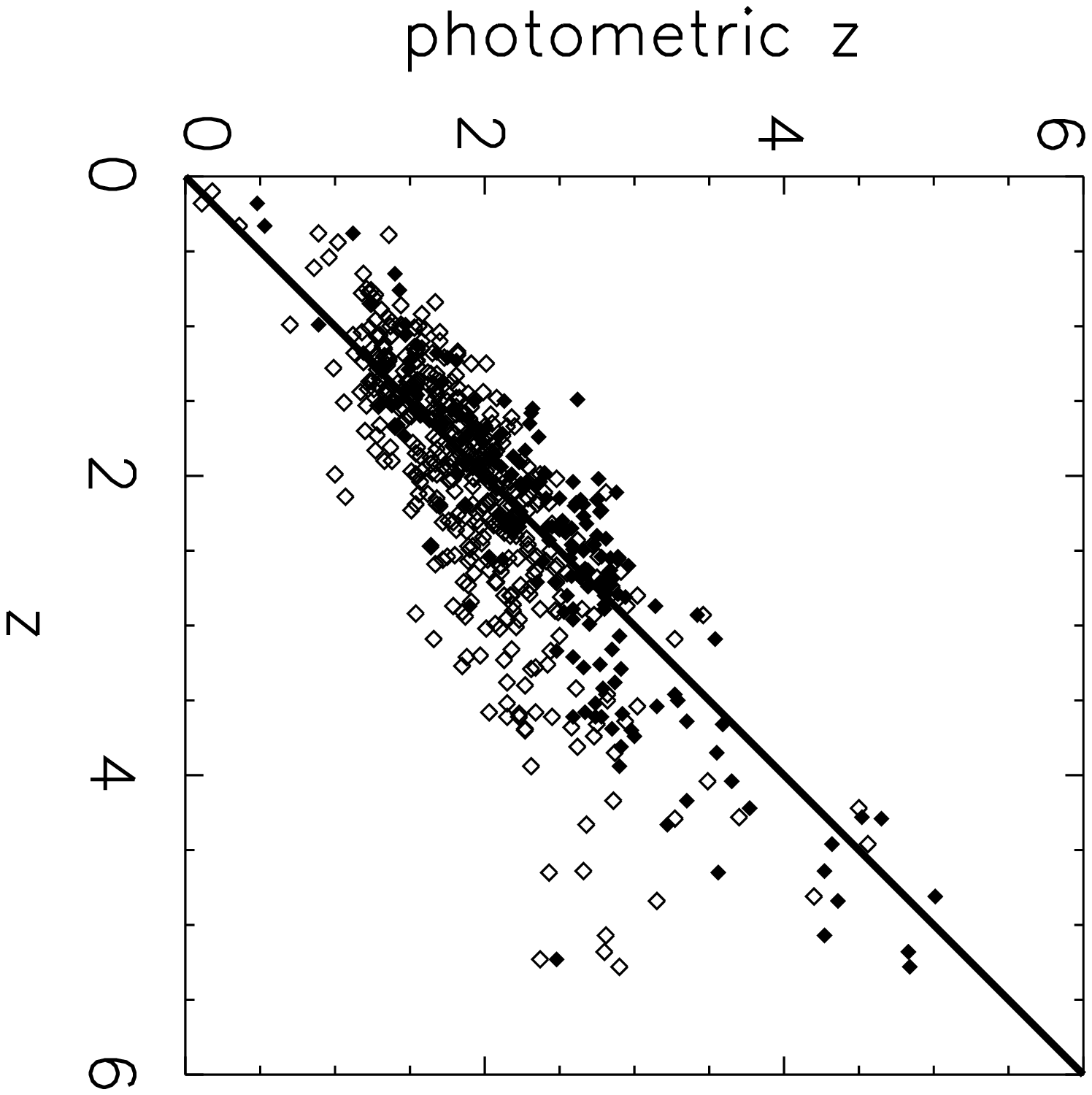}
\includegraphics{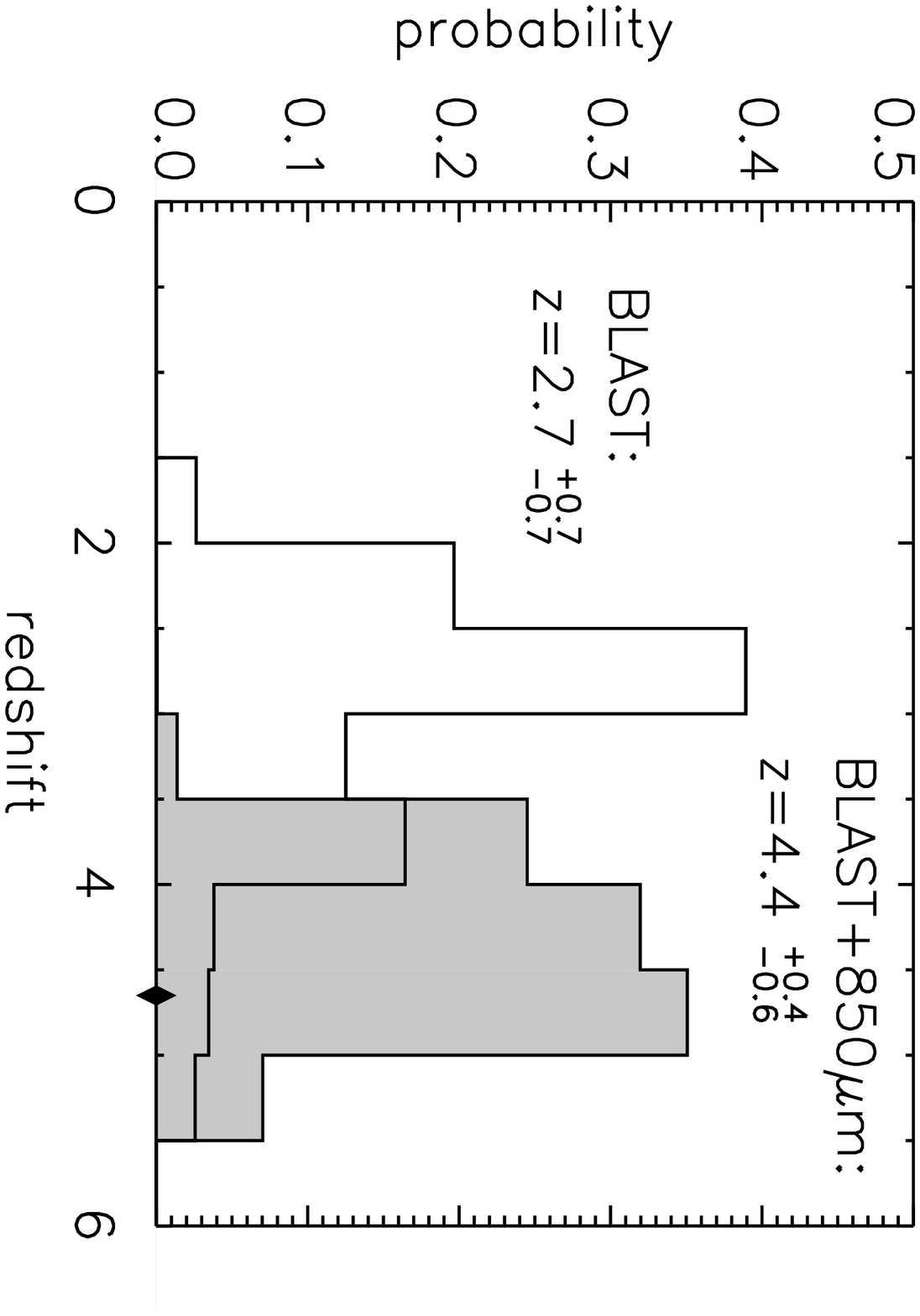} 
\caption{ {\it Left panel:} Photometric redshift vs. true redshift
relationship for 424 mock galaxies simultaneously detected at 250, 350 and
500$\mu$m with {\em BLAST} in a 1~deg$^{2}$ survey. Open symbols show
the relationship inferred using only {\em BLAST} data to derive the
redshifts. Filled symbols show the relationships when the redshifts
are estimated using colours based on {\em BLAST} and complementary
850$\mu$m detections from a ground-based survey. The addition of
850$\mu$m measurements significantly increases the accuracy at $z \geq
4$ since, at these redshifts, {\em BLAST} filters sample the
rest-frame mid-IR to FIR ($\sim 35-100\mu$m). The longer wavelength
data bracket the rest-frame FIR peak of the highest-$z$ objects, which
enhances the diagnostic power for the photometric technique discussed
in this paper.  {\it Right panel:} Example of the correction obtained
for galaxies at $z > 4$ when 850$\mu$m observations are included in
the photometric redshift analysis.  The redshift distributions for a
$1 \times 10^{13} L_{\odot}$ galaxy at $z = 4.65$ with detections at
250$\mu$m (18\,mJy), 350$\mu$m (29\,mJy), 500$\mu$m (27\,mJy) and
850$\mu$m (20\,mJy) are shown as open ({\em BLAST} detections only)
and shaded ({\em BLAST} and 850$\mu$m detections) histograms. This
correction, whilst significant, only applies to 4 per cent of the entire
sample detected in 3 bands.}
\label{fig:BLAST_SCUBAzz}
\end{figure*}

\subsubsection{{\em BLAST} colours}
Fig.\ref{fig:BLASTz} shows two examples of the redshift distributions
for sub-mm galaxies detected by {\em BLAST} at different levels of
significance in all 3 filters.  A further example of the broader
distribution derived for a higher-$z$ galaxy detected only at
500$\mu$m is also shown in Fig.\ref{fig:BLASTz}. The accuracy of the
redshifts determined from {\em BLAST} data is described in more detail
in \S2.4, where we also compare the accuracy to those obtained from
optical--IR colours and the 850$\mu$m/1.4\,GHz spectral index. The
dispersion in the sub-mm photometric redshifts derived from 3 {\em
BLAST} bands is $\Delta z = \pm 0.6$ averaged over all redshifts in
the range $0 < z < 6$.

\subsubsection{SPIRE colours}
An increase in the sensitivity of the sub-mm observations (with SPIRE
for instance) naturally translates into an increased accuracy of the
redshift distributions.  Fig.\ref{fig:SPIRE_colours} shows the
increased sample-size (at all redshifts) for the deeper SPIRE survey
(compared to Fig.\ref{fig:BLAST_colours}). However the greater
sensitivity of the SPIRE observations ($1\sigma = 2.5$\,mJy) does not
significantly improve the accuracy of the redshift distributions,
compared to those determined from detections in all 3 {\em BLAST}
filters. The redshift uncertainties for these sources are due to a
combination of the absolute calibration errors and the dispersion in
the template SEDs.  However SPIRE does provide a far greater redshift
accuracy for the sources that {\em BLAST} detects in only one or two
filters with low S/N (Fig.\ref{fig:SPIREz}). Thus, if the advantage of
the larger aperture of Herschel, the reduced confusion limit and
higher sensitivity of SPIRE is to be realised, then a clearer
understanding of the SEDs of high-$z$ galaxies is first required.

\subsubsection{{\em BLAST} + 850$\mu$m colours}
A greater improvement in the accuracy of the sub-mm photometric
redshifts can be achieved by following-up {\em BLAST} (or SPIRE) surveys
with longer-wavelength survey data (e.g.  ground-based 850$\mu$m
SCUBA data), or vice-versa.  Fig.\ref{fig:BLAST_SCUBAz} shows the
result of simulations that combine {\em BLAST} and 850$\mu$m
observations. The SCUBA data include a photometric measurement error
of 2.5\,mJy, to compare with the shallow UK 8mJy SCUBA survey
\cite{scott01}, and an absolute calibration error of 9 per cent. The addition
of a longer-wavelength datum to the colour information has two
advantages. First, the inclusion of one more colour simply assists the
ability to constrain the redshift. This is particularly important for
galaxies detected only at 500 and/or 350$\mu$m (40 per cent of the sources in
the {\em BLAST} surveys), which lie at the highest redshifts.  Second, at $z
> 4$ the {\em BLAST} filters all begin to sample the short-wavelength
(rest-frame mid-IR) side of the FIR peak.  This 850$\mu$m extension to
the {\em BLAST} wavelength coverage (250--500$\mu$m) ensures that there is
still a long-wavelength datum that brackets the rest-frame FIR peak
(at least until $z \sim 7$).  A clear illustration of these benefits
is shown in Fig.\ref{fig:BLAST_SCUBAzz} where high-$z$ galaxies at $z
> 4$, that had previously been confused with lower-$z$ galaxies at $z
= 2 - 3$ (using only data at 250--500$\mu$m), are now symmetrically
distributed about the $z = z_{\rm phot}$ regression line over the
entire range $0 < z < 6$, with an improved average error of $\Delta z
\sim \pm 0.4$. This correction applies only to a small fraction (4 per cent)
of the galaxies detected in all 3 {\em BLAST} bands. This sub-mm method
therefore continues to provide un-biased estimates of photometric
redshifts for the most distant galaxies.

Given this benefit of including longer-wavelength data, the deepest
{\sl BLAST} surveys will be selected to overlap the largest
(0.5--1.0~deg$^{2}$) surveys carried out by the JCMT at 850~$\mu$m and
IRAM at 1.25\,mm.  The extended wavelength coverage (250$\mu$m --
1.1\,mm) will provide accurate redshift constraints out $z \sim 8$.

The analysis described in this paper simultaneously makes use of all
of the available multi-frequency data (unlike the simple one--colour vs.
redshift treatments described elsewhere), and takes into account the
measurement errors and uncertainties in the SEDs of the template
galaxies used to extrapolate the rest-frame luminosities of the
sub-mm galaxies from their observed data.  

\begin{figure}
\vspace{18.5cm} 
\includegraphics{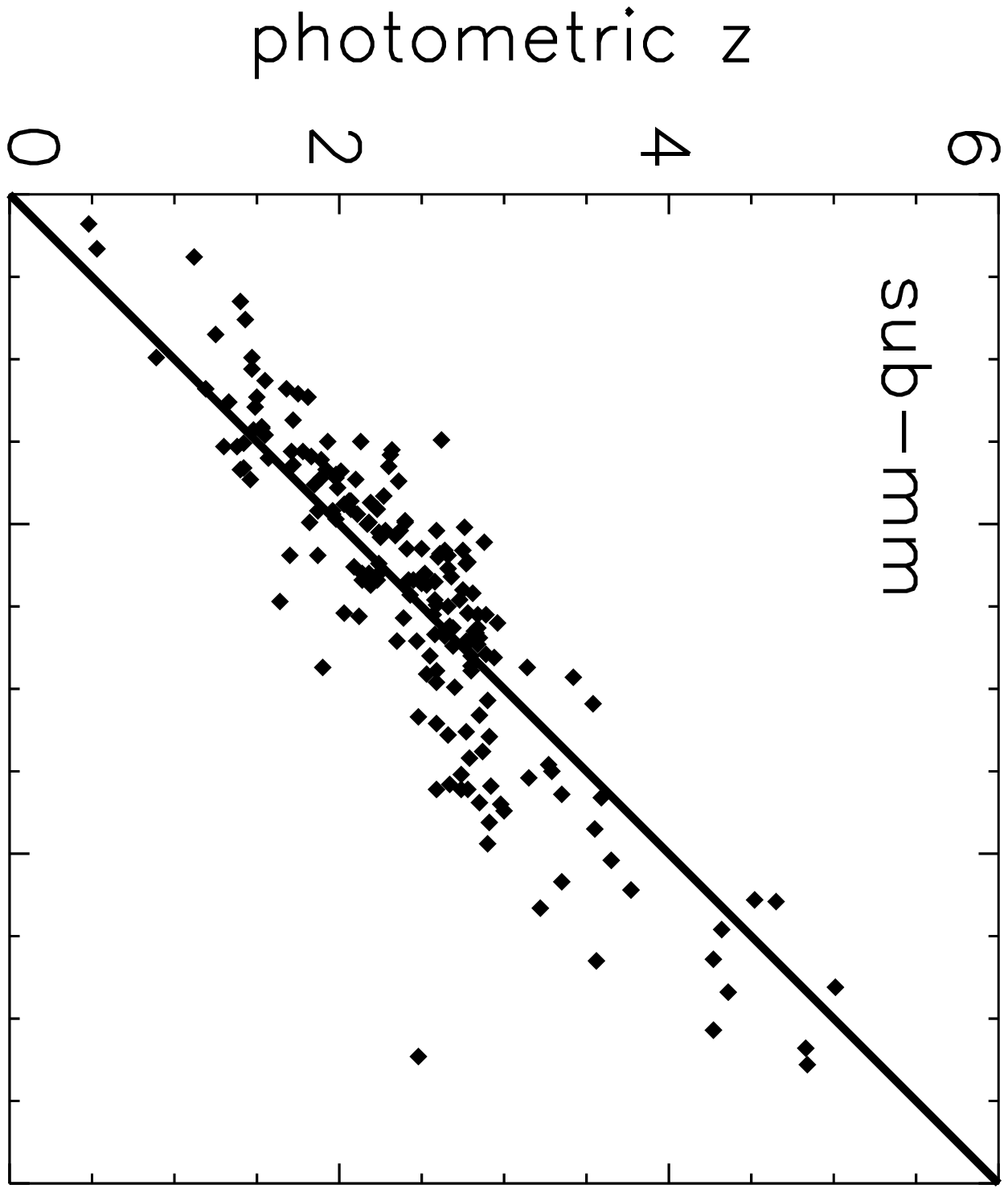} 
\includegraphics{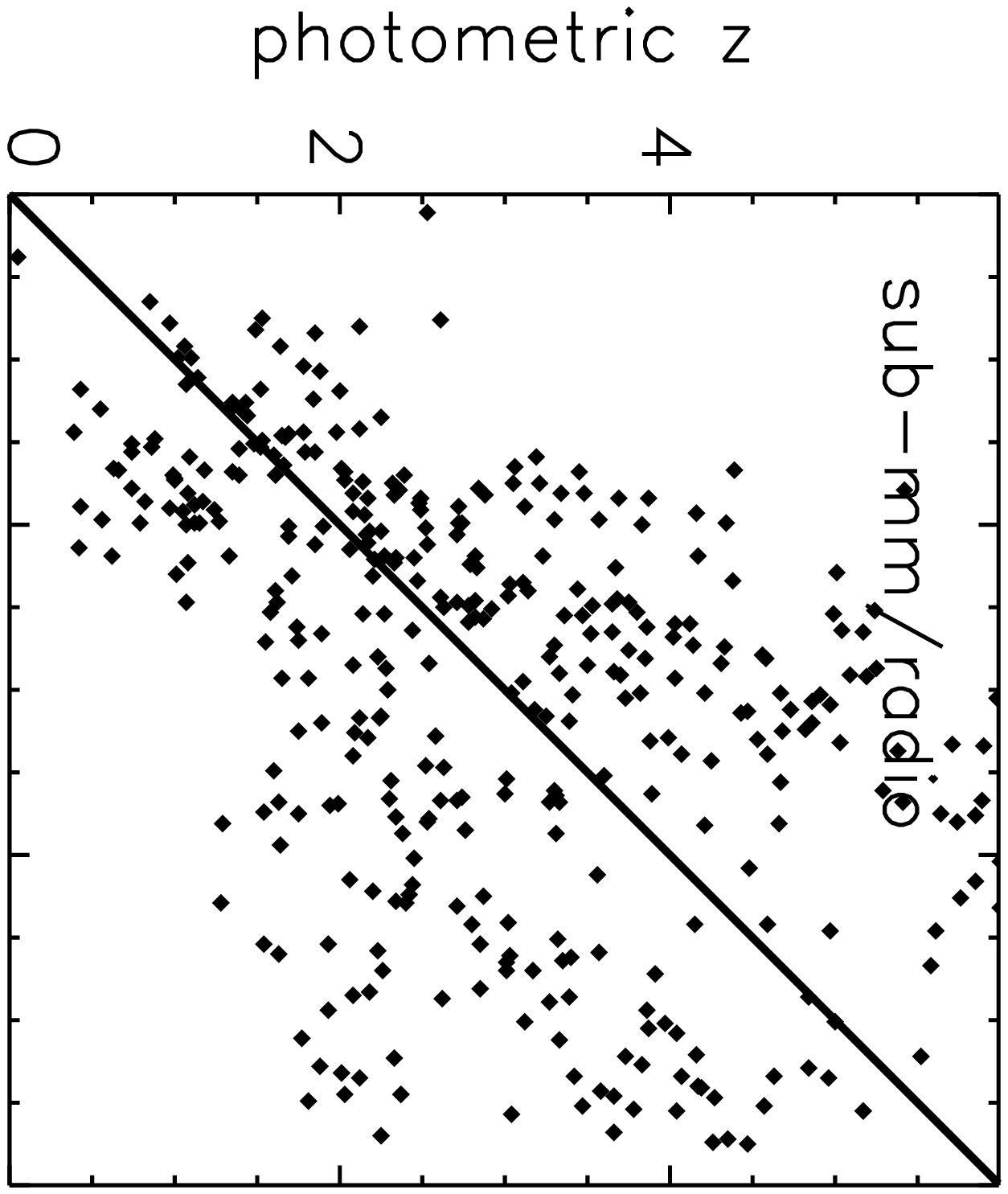} 
\includegraphics{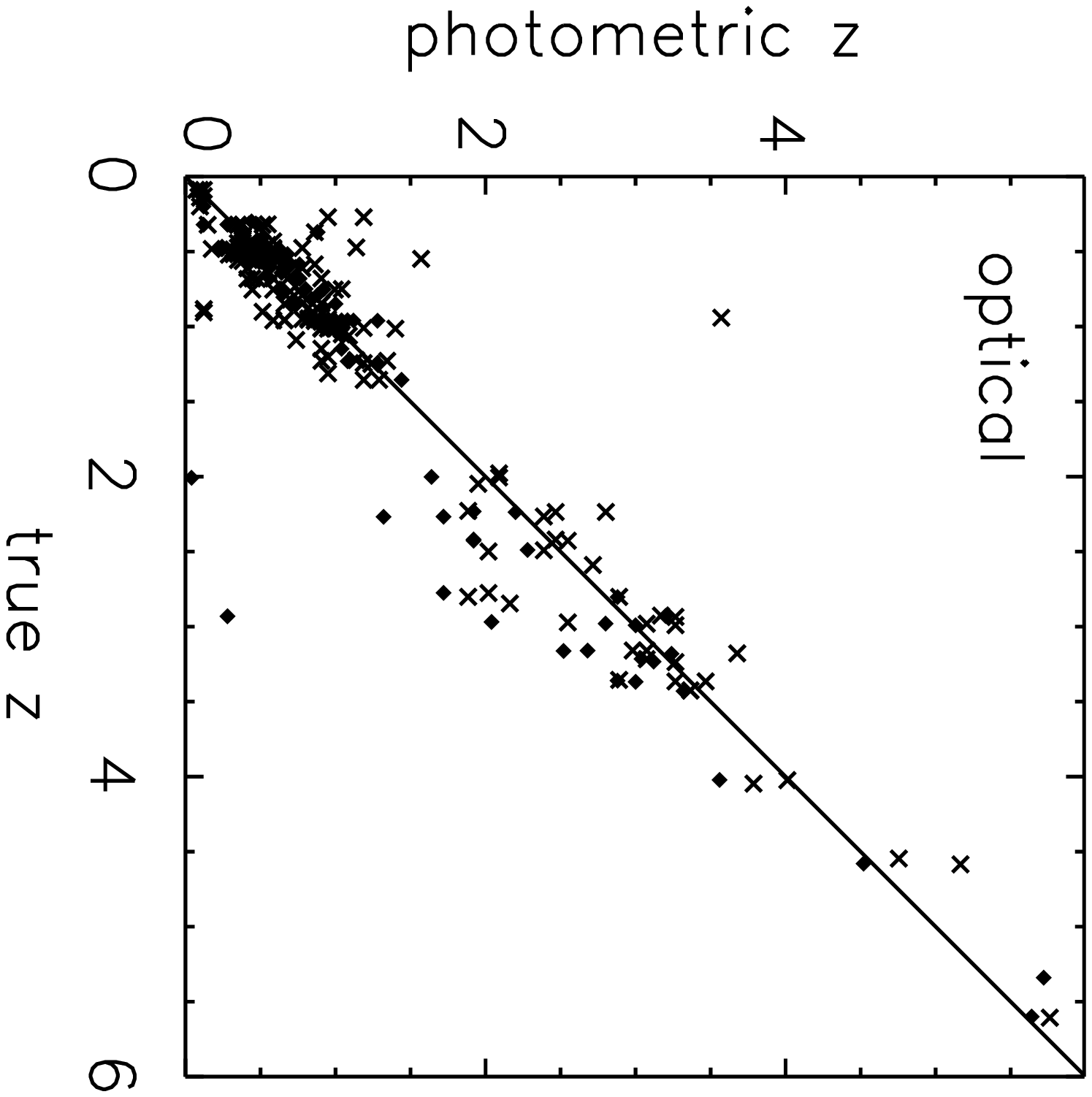} 
\caption{{\it Top panel:} Sub-mm photometric redshifts vs. true
redshifts for 189 mock galaxies simultaneously detected at 250, 350,
500 and 850$\mu$m in 1~deg$^{2}$. The overall r.m.s. scatter in the
sub-mm photometric redshifts, determined from the combined {\em BLAST} and
SCUBA 850$\mu$m data, is $\Delta\,z ({\rm submm}) \sim 0.4$ in the
redshift range $0 < z < 6$; {\it Middle panel:} Sub-mm--radio
photometric redshifts vs. true redshifts for galaxies selected from
the same catalogue as in the top panel, using $\alpha^{350}_{1.4}$,
the 350--1.4\,GHz spectral index described by Carilli \& Yun (2000);
{\it Bottom panel:} Optical--IR photometric redshifts vs.
spectroscopic redshifts in the Hubble Deep Field - 108 galaxies
(Fern\'andez-Soto, Lanzetta, Yahil 1999 - diamonds), 156 galaxies
(Rowan-Robinson 2001 - crosses).  The optical--IR photometric
redshifts, for galaxies at $z_{\rm true} \geq 2$, show a dispersion of
$\Delta\,z ({\rm opt}) \sim 0.5$ (excluding the outlying sources).}
\label{fig:zz_comparison}
\end{figure}

\subsection{Comparison of Optical--IR, Sub-mm and Sub-mm--Radio Photometric Redshifts}

The photometric redshifts of galaxies derived from sub-mm {\em BLAST}
data (250, 350 and 500$\mu$m), using the method described above, are
compared with the true redshifts in our mock catalogues. We have
demonstrated that an average accuracy of $\Delta\,z ({\rm submm}) \sim
0.6$ can be achieved over the range $0 < z < 6$, although, as Table~2
shows, the scatter increases significantly for $z > 3.5$ for the
reasons described in \S2.3.  The photometric redshifts measured from
the optical-IR SEDs of galaxies in the HDF \cite{soto99,mrr} have an
accuracy that varies between $\Delta\,z ({\rm opt}) \sim 0.1 - 0.6$
over the same redshift range.  However we note that the optical
galaxies described in this HDF analysis are obviously not drawn from
the same population as the sub-mm galaxies. The accuracy of the
optical photometric redshifts must be much poorer if the technique is
applied to the optically-obscured SCUBA galaxies.  The inclusion of
longer wavelength data (e.g. 850$\mu$m) to the {\em BLAST} data
improves the photometric redshift accuracy to $\Delta\,z ({\rm submm})
\sim 0.4$, and extends the redshift range over which we can measure
reliable redshifts to $0 < z < 6$ (Fig.\ref{fig:zz_comparison} and
Table~2).  This means that, despite the greater wealth of data
describing the optical--IR SEDs of galaxies and the number of
available optical--IR bands, the sub-mm photometric redshifts of the
sub-mm galaxy population can be determined with a similar accuracy to
the optical-IR photometric redshifts of optically-selected galaxies at
$z > 1.5$ (Fig.\ref{fig:zz_comparison}).  

We also draw attention to the lowest-redshift bin, $0.5 < z < 1.5$, in
Table~2, which shows that the addition of 850$\mu$m data decreases the
redshift accuracy, compared to that derived from the shorter
wavelength 250--500$\mu$m observations. This is due to the fact that
the brightness of low-redshift galaxies, assuming our evolutionary
model for the luminosity function, will necessarily have low S/N
detections in an 850$\mu$m survey with a $1\sigma$ sensitivity of
2.5\,mJy.  Thus the low S/N 850$\mu$m detections, when combined with
the BLAST data, allows a greater range of redshifts to be consistent
with the observed colours.  Conversely the addition of shorter
wavelength SIRTF data with high S/N will improve the accuracy of our
photometric redshift estimates at $z < 1.5$.

The redshift accuracies determined from {\em BLAST} and {\em
BLAST}+850$\mu$m data (cols. 3 and 4 in Table~\ref{tab:redshifts}) are
compared with those obtained from the identical data in our mock
catalogues using the 350\,GHz (850$\mu$m) to 1.4\,GHz spectral index
($\alpha^{350}_{1.4}$) described by Carilli \& Yun (2000). The
simulations include a photometric error of 7$\mu$Jy, and absolute
calibration error of 3 per cent to compare with the deepest VLA surveys. As
Fig.\ref{fig:zz_comparison} clearly illustrates, the photometric
redshifts obtained with the sub-mm/radio method suffer a large scatter
($\Delta\,z (\alpha^{350}_{1.4}) > 1.5$), particularly at $z > 2$,
where small differences in $\alpha^{350}_{1.4}$ result in large
differences in the redshift estimates. The inability of the
850--1.4\,GHz index to determine accurate redshifts at $z > 2$ has
already been noted by Carilli \& Yun \shortcite{carilliyun00}.  At $z
< 2.5$, photometric redshifts using $\alpha^{350}_{1.4}$ are still
less accurate than those determined from the sub-mm (250--850$\mu$m)
method described in this paper. This is due, in part, to the larger
scatter of the SEDs at radio wavelengths
(Fig.\ref{fig:seds1}). Naturally, any redshift estimated from a
single colour index (e.g. $\alpha^{350}_{1.4}$) is also, in
general, going to a offer a less stringent constraint than the use of
two or more colours.

\begin{table}
\begin{tabular}{ccccc}
\hline redshift bin & \multicolumn{4}{c}{redshift accuracy $\Delta z$}\\ 
z & optical & {\em BLAST} & {\em BLAST} & 850$\mu$m\\ 
  & + IR    &             & + 850$\mu$m & + 1.4\,GHz\\ \hline 
0.5~$-$~1.5 & 0.1 (0.2) & 0.3 & 0.5 & 0.7 \\
1.5~$-$~2.5 & 0.5 (0.3) & 0.3 & 0.3 & 1.2 \\ 
2.5~$-$~3.5 & 0.5 (0.4) & 0.7 & 0.3 & 1.4 \\ 
3.5~$-$~4.5 & 0.5 (0.2) & 1.1 & 0.7 & 2.1 \\ 
4.5~$-$~5.5 & 0.3 (0.6) & 2.1 & 0.6 & 2.5 \\ \hline
\end{tabular}

\caption{Comparison of the $1\sigma$ redshift accuracy, $\Delta\,z$,
determined from a simulated survey with r.m.s sensitivities of 5\,mJy
at 250, 350 and 500$\mu$m, 2.5\,mJy at 850$\mu$m and 7$\mu$Jy at 1.4GHz
using different photometric colours in redshift bins between $0.5 < z
< 5.5$: Column\,2 - optical data (Fernandez-Soto et al. 1999, and
Rowan-Robinson 2001 in parentheses); column\,3 - sub-mm data from
{\em BLAST} at 250, 350 and 500$\mu$m; Column\,4 - extended sub-mm data from
{\em BLAST} and additional 850$\mu$m observations; Column\,5 - sub-mm
(850$\mu$m, 350\,GHz) and radio (1.4\,GHz) data.  Redshifts have been
calculated from $\alpha^{350}_{1.4}$, defined as the spectral index
between 350 and 1.4\,GHz, using the formalism described by Carilli \&
Yun (2000).  The redshifts in Cols.\,3, 4 and 5 have been calculated
for galaxies drawn from the same mock catalogue.}
\label{tab:redshifts}
\end{table}

\begin{figure*}
\vspace{9cm} 
\includegraphics{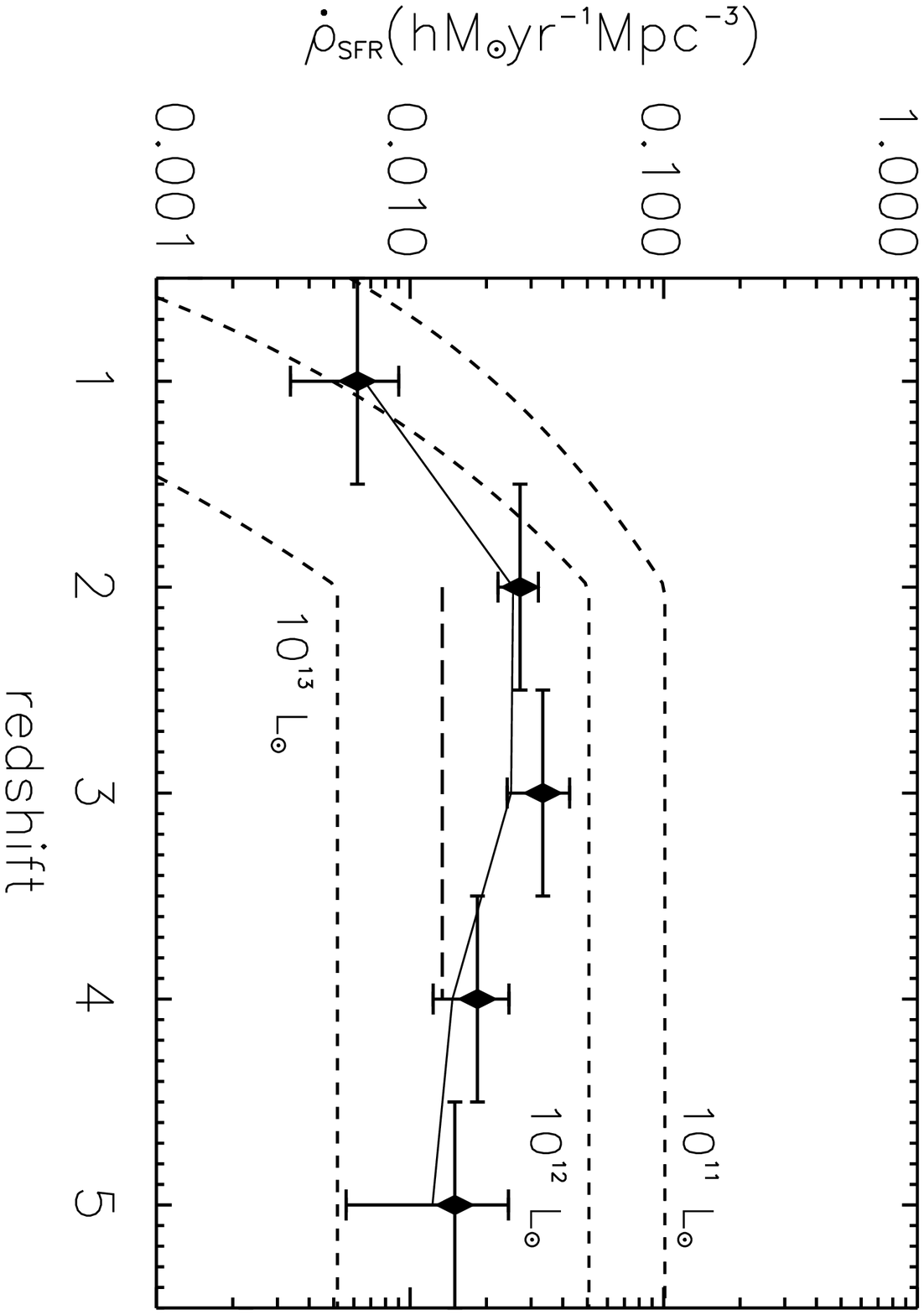} 
\includegraphics{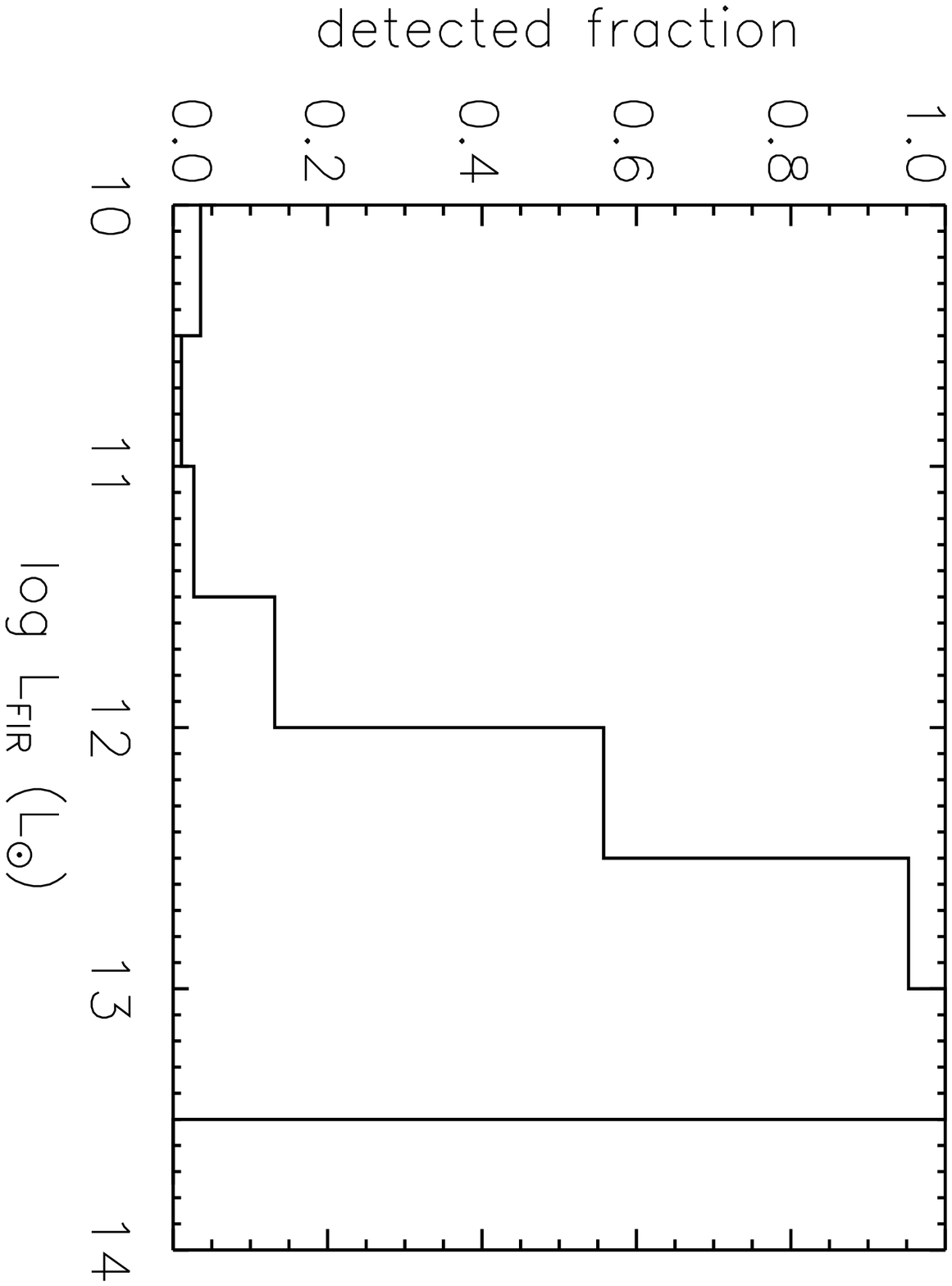}
  \caption{{\em Left panel}: Global star formation rate history of
  galaxies detected in at least 2 filters from a combined {\em BLAST}
  (3$\sigma=15$~mJy) and wide-area SCUBA 850$\mu$m survey
  (3$\sigma=8$~mJy). Filled diamonds represent the recovered values of
  the SFR using the photometric redshift distributions. We compare
  these values with the true SFR included in the mock catalogue (shown
  as a solid line) before measurement errors are introduced.  The
  $3\sigma$ error bars of the reconstructed SFR are dominated by the
  uncertainty of the SED that should be used to derive the 60$\mu$m
  luminosities of the mock galaxies (which have intrinsically
  different SEDs), and not by the accuracy of the redshift
  determinations.  The long-dashed horizontal line shows an estimate
  of the minimum SFR density for sub-mm galaxies that have been
  detected in the UK 8mJy SCUBA 850$\mu$m survey (Scott et al. 2001),
  assuming that the SCUBA galaxies lie between $2 < z < 4$.  The
  short-dashed curves represent the theoretical SFR due to galaxies of
  FIR luminosities $>10^{11-13} L_{\odot}$ derived from the
  evolutionary model adopted in the simulations \S2.2.  {\em Right
  Panel}: Fraction of galaxies, per luminosity interval, detected at
  all redshifts in the mock combined {\em BLAST} and wide-area SCUBA
  survey.  The completeness of the survey is $\sim$98 per cent for galaxies
  $> 3 \times 10^{12} L_{\odot}$.}
\label{fig:sfr_history}
\end{figure*}

\subsection{Measuring the Star Formation History from Sub-mm Surveys}

The measurement of redshift probability distributions for a large
sample of sub-mm galaxies provides the information necessary to derive
robust estimates of global properties of the population, such as the
SFR history, even if the redshift and luminosity of individual
galaxies are not known with a great degree of precision.  The
photometric redshift distribution of an individual galaxy allows the
determination of its probable contribution to the total FIR luminosity
density in a given redshift interval.

This calculation is performed via 100 Monte-Carlo samples drawn from
the redshift probability distribution for each galaxy in the mock
survey.  The FIR luminosities of these galaxies are then derived from
the fluxes in the mock catalogue and the new synthetic redshifts.  The
SED from the template library that best represents the colours
(including observational errors) of the whole population is used to
calculate the necessary k-correction. This SED will in general be
different from that assigned in the original catalogue.  We then
calculate the uncertainty in the k-correction, and therefore the error
in the FIR luminosity, by randomly selecting any SED in the template
library.  Thus we remove any bias introduced into the reconstruction
of the SFR through the choice of a single SED intended to represent
the whole population of sub-mm galaxies. Finally, the star formation
density is calculated from the mean of the Monte-Carlo FIR
luminosities, assuming ${\rm SFR}/M_{\odot} {\rm yr}^{-1} = 2.0 \times
10^{-10} L_{\rm FIR}/L_{\odot}$ \cite{hughes97}.  

A deep {\em BLAST} survey, combined with a wide-area shallow SCUBA
survey ($S_{850\mu{\rm m}} > 8$~mJy), can recover the SFR density due
to galaxies with $L_{\rm FIR} \ge 3 \times 10^{12} L_{\odot}$ at
$0.5\le z\le 5.5$ with an accuracy of $\sim 20$ per cent
(Fig.\ref{fig:sfr_history}).  The completeness of the survey is $\sim
98$ per cent above $3 \times 10^{12} L_{\odot}$ (Fig.\ref{fig:sfr_history}).
The SFR density derived from the mock combined {\em BLAST} and SCUBA
survey is consistent with the preliminary results from the UK 8~mJy
SCUBA survey \cite{scott01}, and provides a good illustration of the
advantage of being able to derive photometric redshifts for the entire
survey sample over a large redshift range.

\section{Conclusions} 
Monte-Carlo simulations of multi-wavelength (250--850$\mu$m) sub-mm
surveys show that the majority of the scatter in the derived sub-mm
colours will be due, in approximately equal parts, to the dispersion
in the SEDs of galaxies in a library of templates, and the
observational and calibration errors. Since we can control to some
extent the quality of the experimental data, it is still the
uncertainty in the observed (or model) SEDs of high-$z$ starburst
galaxies that will ultimately limit the accuracy of these photometric
redshift predictions.  Theoretical SEDs, based on radiative transfer
models for high-$z$ starburst galaxies \cite{efstathiou}, can help in
principle.  However the low S/N, and restricted wavelength coverage of
the available observational data for high-$z$ starburst galaxies are
consistent with such a broad range of theoretical models that they
currently provide little discriminatory power in determining the most
appropriate SEDs to use as templates.  Consequently, one of the most
straight-forward achievements of {\em BLAST}, but no less important,
will be simply to provide an accurate empirical model of the
temperature-sensitive SEDs for 1000's of high-$z$ galaxies at these
critical short sub-mm wavelengths (250--500$\mu$m).

The product of the simulations of sub-mm surveys described in this
paper is the ability to calculate the probability distribution for the
redshift of any individual galaxy, taking into account observational
errors and the uncertainty in the appropriate template SED, without
the requirement to first identify the optical, IR or radio
counterpart.  These simulations demonstrate that the combination of
balloon-borne (airborne or satellite) short-wavelength sub-mm data at
250--500$\mu$m, and longer-wavelength ground-based 850$\mu$m data, for
a statistical sample of galaxies can provide the rest-frame FIR
luminosity distribution, and hence the star formation history of the
entire sub-mm population.

The precision of redshift estimates determined from short-wavelength
sub-mm observations (250--500$\mu$m) is encouraging, with a $1\sigma$
dispersion of $\Delta\,z ({\rm submm}) = 0.4 - 0.6$ (depending on the
availability of longer-wavelength ground-based data, $\lambda \geq
850\mu$m) over the range $0 < z < 6$. This sub-mm precision is
comparable to the accuracy of the optical-IR photometric estimates of
redshift, $\Delta\,z ({\rm opt}) \sim 0.5$ at $z > 2$, and
significantly better than those provided by the 850$\mu$m/1.4\,GHz
spectral index.  An important consequence of these
sub-mm redshift constraints is the ability to measure the
star-formation of luminous sub-mm galaxies ($L_{\rm FIR} > 3 \times
10^{12} L_{\odot}$) with an unprecedented error of $\sim 20$ per cent.

The advantage of having a constrained redshift probability
distribution for individual sub-mm galaxies, without optical, IR or
radio (continuum) counterparts, is that we can now determine the
likelihood that a redshifted rotational CO transition-line falls into
the frequency range of any particular spectral-line receiver on the
next generation of large mm--cm telescopes (e.g. 100-m GBT, 50-m
LMT).  The opportunities to conduct mm-wavelength spectroscopic
follow-up observations that will provide definitive molecular-line
redshifts, and dynamical-mass estimates of sub-mm galaxies, are discussed
further in Paper II \cite{aretxaga02}.

The combination of short-wavelength sub-mm data from {\em BLAST},     
SPIRE or other facilities, and large single-dish millimetre wavelength
telescopes provides a powerful combination to break the {\em redshift 
deadlock} that hinders our ability to understand the evolution and
nature of the sub-mm starburst galaxy population.

\section{Acknowledgments}
We thank the anonymous referee for constructive suggestions to improve 
this paper.
This work has been funded by CONACYT grants 32180-E and 32143-E.
The development of {\em BLAST} is supported in part by NASA grant
NAG5-92291. DHH, IA, and ELC would like to thank Mark Devlin and the
Physics \& Astronomy Dept. at Univ. of Pennsylvania,
Philadelphia for their financial support and hospitality in June 2001
during which time part of this work was carried out.

\end{document}